\documentclass{aa}
\usepackage{epsfig}
\usepackage{lscape}
\usepackage{aalongtable}
\usepackage{natbib}
\usepackage{graphics}
\usepackage{txfonts}

\begin{document}

\def\h2{H {\sc ii}}

\title{Balmer jump temperature determination in a large sample of 
low-metallicity H {\sc ii} regions\thanks{Based on observations
collected at the European Southern Observatory, Chile, ESO program 
76.B-0739.}$^,$\thanks{Tables 3 and 4 are only available in electronic form
at the CDS via anonymous ftp to cdsarc.u-strasbg.fr (130.79.128.5)
or via http://cdsweb.u-strasbg.fr/cgi-bin/qcat?J/A+A/}
}

\author{ N. G. Guseva \inst{1}
\and Y. I.\ Izotov \inst{1}
\and P.\ Papaderos \inst{2,3}
\and K. J.\ Fricke \inst{3}}
\offprints{N. G. Guseva, guseva@mao.kiev.ua}
\institute{      Main Astronomical Observatory,
                     Ukrainian National Academy of Sciences,
                     Zabolotnoho 27, Kyiv 03680,  Ukraine
\and
{Instituto de Astrof\'{\i}sica de Andaluc\'{\i}a, Apdo. 3004, 18080 Granada, Spain}
\and                    
                     Institute for Astrophysics, Friedrich-Hund-Platz 1,
                     37077 G\"ottingen, Germany
}

\date{Received \hskip 2cm; Accepted}

\abstract{}{Continuing the systematic determination of the electron temperature of \h2\ regions
using the Balmer and/or Paschen discontinuities by  Guseva et al. (2006) we focus here on 3.6m ESO 
telescope 
observations of a large new sample of 69 \h2\ regions in 45 blue compact dwarf (BCD) 
galaxies. 
This data set spans a wide range in metallicity ($Z_{\odot}/60 \la Z \la Z_{\odot}/3$) 
and, combined with the sample of 47 \h2\ regions from  Guseva et al. (2006), yields the largest 
spectroscopic data set ever used to derive the electron temperature in the H$^+$ zone.}
{In the same way as in Guseva et al. (2006) we have used a Monte Carlo technique 
to vary free parameters and to calculate a series 
of model spectral energy distributions (SEDs) for each \h2\ region.
The electron temperature in the H$^+$ zones was derived from the best fitting synthetic 
and observed SEDs
in the wavelength range $\sim$ 3200--5100\AA, which includes the Balmer jump.}
{On the base of the present large spectroscopic sample
we find that in hot ($T_e$(H$^+$) $\ga$ 11000 K) \h2\ regions the temperature 
of the O$^{2+}$ zone, determined from doubly ionised oxygen forbidden lines, does not 
differ statistically from the temperature of the H$^+$ zone.
Thus, we confirm and strengthen the finding by Guseva et al. (2006).
We emphasize that due to a number of modelling assumptions
and the observational uncertainties for individual objects,
only a large, homogeneous sample, as the one used here, 
can enable a conclusive study of the relation between $T_e$(H$^+$) and
$T_e$(O {\sc iii}).}{}
\titlerunning{Balmer jump temperature in low-metallicity H {\sc ii} regions}
\maketitle

\section {Introduction}
\label{intro}
 A long-standing problem in the study of H {\sc ii} regions is connected to 
the fact that the electron temperatures derived from the hydrogen Balmer 
and Paschen discontinuity or from 
the UV helium discontinuity are systematically lower than those derived from the 
collisionally excited optical [O {\sc iii}] lines which are most often 
used for the temperature determination \citep[e.g. see review in ][]{O2003}.

  Electron temperature determinations based both on collisionally 
excited line diagnostics and on the Balmer and Paschen jumps
have been mainly applied to planetary nebulae (PNe) 
\citep[e.g. ][]{Liu2000,Luo2001,R03,Z2004,P04,W05}
and to nearly solar-metallicity  H {\sc ii} regions 
in the Milky Way and several nearby galaxies with relatively high 
metallicities of 1/3 -- 1/10 solar 
\citep[e.g. ][]{P92,P93,P00,P03,E98,G04,G05,G06,G_D94}.
A common result of these studies is that, generally, $T_e$(H$^+$) is lower
than  $T_e$(O {\sc iii}). Only recently  \citet{GIT2006}  have derived the electron 
temperature of a relatively large sample of low-metallicity 
H {\sc ii} regions using  the Balmer and Paschen jumps.

Differences between $T_e$(H$^+$) 
and $T_e$(O {\sc iii}) were first discussed by
\citet{P67}, who introduced a concept of temperature fluctuations in the nebulae.
To quantify temperature fluctuations he  used the parameter $t^2$, the 
mean square temperature variation. It was shown in many studies 
of planetary nebulae and H {\sc ii} regions
\citep[e.g. ][]{Peimb95,E98,E99,E2002,Liu2000,Liu2001}
that the heavy element abundances derived from the recombination lines and
from the collisionally excited lines are consistent if $t^2$ is in the
range 0.02 -- 0.10. On the other hand, typical values of $t^2$
for photoionization models of chemically and spatially homogeneous 
nebulae are significantly lower, 
$t^2$ = 0.00 -- 0.02 \citep[e.g. ][]{Kingdon1995,Perez97}.

In recent years point-to-point measurements of the electron 
temperature fluctuations have been done in several PNe and H {\sc ii} regions 
\citep{Liu1998,Rubin2002,KC05,Rubin2003,O2003}.
Only low amplitude temperature variations were found across 
the nebulae. 
 This is probably because the projected parameter $t_s^2$ 
derived in the point-to-point measurements of the electron temperature
is different from the total parameter $t^2$. 
 \citet{Cop06} has shown that  $t_s^2$ gives only a low limit of $t^2$.
 Additionally, there are many difficulties to obtain the temperature maps of 
the whole nebulae when images and spectra obtained with different space-born 
and ground-based telescopes are combined \citep{Lur03}.

Thus, the observational data for PNe and H {\sc ii} regions with 
relatively low electron temperatures $\leq$11000K certainly show  
systematic differences between the temperatures obtained from the 
collisionally excited lines and 
from the Balmer and Paschen jumps 
\citep[see Fig.12 in ][for the data collected from literature]{GIT2006}.
On the other hand, no significant differences 
between $T_e$(H$^+$) and $T_e$(O {\sc iii}) were found for the 
higher-temperature H {\sc ii} regions \citep{GIT2006}.
This conclusion is in agreement with the results obtained by
\citet{PPLurid2002}, who compared $T_e$(O {\sc iii}) and the electron
temperature $T_e$(He$^+$) in the He$^+$ zone. 
Adopting $t^2$ = 0.01 -- 0.04, these authors 
found the temperature differences 
of (1 -- 3)\% in H {\sc ii} regions with $T_e$(O {\sc iii}) = 20000K
and of (3 -- 12)\% in H {\sc ii} regions with $T_e$(O {\sc iii}) = 10000K. 
 As $T_e$(He$^+$) $<$ $T_e$(H$^+$) in PNe
\citep{Z05} and $T_e$(He$^+$) $\approx$ $T_e$(H$^+$) in 
H {\sc ii} regions \citep{G05}, the differences between $T_e$(O {\sc iii}) and $T_e$(H$^+$)
are expected to be not greater than (1 -- 3)\% in H {\sc ii} regions with 
the temperatures $T_e$(O {\sc iii}) $\sim$ 20000K.

Different mechanisms are proposed to
explain the observational differences in $T_e$. In particular, density and 
chemical abundance inhomogeneities, 
ionization of nebulae by low-energy cosmic rays are considered
\citep[e.g. ][]{T_P90,V94,Liu2000,T04,TP05,Giam05}. However, no definite
conclusions have been made concerning the main mechanism 
for these differences.

The knowledge of differences in $T_e$ is important for the element
abundance determination.
  The heavy element abundances are usually derived assuming that
the temperature of the H$^+$ zone is equal to the temperature of the O$^{2+}$
zone. If instead, $T_e$(H$^+$) is smaller than $T_e$(O {\sc iii}), then  
the heavy element abundances would be increased. 
  For the determination of 
the primordial He abundance from spectra of low-metallicity BCDs 
\citep[e.g. ][]{IT04} the knowledge of the temperature structure of 
a H {\sc ii} region is especially important because the high precision
of 1 -- 2 \% is required for the He abundance.
  Thus to estimate the
systematic error of the primordial He abundance it is crucial 
to investigate whether temperature differences are as important in 
low-metallicity BCDs as they are in high-metallicity H {\sc ii} regions and 
PNe.  

 \citet{GIT2006} have determined the electron temperature of H$^+$ zones 
from the Balmer jump in 23 H {\sc ii} regions and from the Paschen jump 
in 24 H {\sc ii} regions
in the metallicity range from 1/3 to 1/60 of solar,
based, respectively, on Multiple Mirror Telescope (MMT) and Sloan Digital Sky Survey
(SDSS) data.
  They used Monte Carlo simulations, varying the electron temperature in 
the H$^+$ zone,
the extinction of the ionised gas and that of the stellar population,
the relative contribution of the ionised gas to the total emission and 
the star formation history  
to fit the spectral energy distribution (SED) of the galaxies
in the large wavelength range, which includes the Balmer and Paschen
discontinuities.
  The best sets of free parameters have been obtained from the
minimization of the deviations between observed and modelled SEDs.
 It was found that the temperatures $T_e$(O {\sc iii}) of the O$^{2+}$ zones 
determined from the 
nebular to auroral line flux ratio of doubly ionised 
oxygen [O {\sc iii}] $\lambda$(4959+5007)/$\lambda$4363
do not differ, in a statistical sense, from the 
temperatures $T_e$(H$^+$) of the H$^+$ zones determined from 
Balmer and Paschen jumps.
  On the other hand, \citet{GIT2006} have emphasized that, 
due to large observational 
uncertainties and modelling assumptions
for individual objects, only a statistical study 
of a large sample of H {\sc ii} regions can allow for definite conclusions.

 Therefore in order to study the $T_e$(O {\sc iii}) vs. $T_e$(H$^+$) 
relation over a large metallicity range,
between $Z_{\odot}/60$ and $Z_{\odot}/3$, and with much improved statistics,
we took new spectra  of H {\sc ii} regions with the 3.6m ESO telescope. 
These data are combined with the ones obtained by \citet{GIT2006}.

In Section \ref{obs}, we describe the observations and data reduction
of the new spectroscopic data. 
 In Section \ref{Results} we compare $T_e$(H$^+$) and $T_e$(O {\sc iii}) 
for the new BCD sample and those from literature. 
  Our conclusions are summarized in Section \ref{conc}.

\section{Observations and data reduction \label{obs}}


The sample observed with the 3.6m ESO telescope consists of 69 low-metallicity 
H {\sc ii} regions in 45 BCDs. They were selected mainly from the literature.
Particularly, two galaxies were selected from the Data Release 2 
of the Six-Degree Field Galaxy Redshift Survey (6dFGRS) \citep{Jones_2005}
and eighteen galaxies were selected from the Data Release 4 of 
the Sloan Digital Sky Survey (SDSS) \citep{Adel_2006}.
The BCDs were chosen to span a large range of oxygen abundance, from about 
1/60 to about 1/3 that of the Sun.
Thus, we can study the dependence of temperature variations with metallicity. 
Since the electron 
temperature of a H {\sc ii} region depends on its metallicity, H {\sc ii} 
regions in our BCD sample span also a large range of electron temperatures.
  The equivalent width of the H$\beta$ line, EW(H$\beta$), is a measure of the relative
contribution of the ionised gas emission to the total light.
The selected galaxies cover a large range of EW(H$\beta$) from 
10\AA\ to 382\AA.

  All spectra were obtained with the  
EFOSC2 (ESO Faint Object Spectrograph and Camera) mounted at the 3.6m ESO 
telescope at La Silla in two observing runs, the
first one, during  April 11 -- 14, 2005 and the second one,
during  October 7 -- 9, 2005. The observing conditions were photometric 
during all nights.
  For the spring observations we used the grism $\#14$ and the grating 
600 gr/mm. The long slit with 1\arcsec$\times$300\arcsec\ was centered 
on the brightest part of each galaxy.
The above instrumental setup gave a wavelength coverage of 
$\lambda$$\lambda$3200--5083, a spectral resolution of $\sim$6.2~\AA\ (FWHM)
and a spatial scale of 0\farcs314 pixel$^{-1}$ along the slit for the used
2$\times$2 pixel binning.

  During the fall observations the grism $\#07$ and the grating 600 gr/mm were used,
resulting in a wavelength coverage of $\lambda$$\lambda$3250--5200.
These observations were carried out with a 1\farcs2$\times$300\arcsec\ slit
centered on the brightest part of each galaxy. The spectral resolution and spatial 
scale along this slit were $\sim$6.2~\AA\ (FWHM) and 0\farcs157 pixel$^{-1}$, 
respectively.

Our sample galaxies were mostly observed at low airmass $<$1.2. 
Spectroscopic observations at a higher airmass were carried out along the parallactic angle.
Thus, no corrections for atmospheric refraction have been applied. 
The total exposure time of typically 40 -- 60 minutes per galaxy was split up 
into 2 -- 3 subexposures  to allow for a more 
efficient rejection of cosmic ray hits.
Three spectrophotometric standard stars were observed during each night
for flux calibration.
The journal of the observations is given in Table~\ref{journal}.

The data reduction was carried out with the IRAF\footnote{IRAF is 
the Image Reduction and Analysis Facility distributed by the 
National Optical Astronomy Observatory, which is operated by the 
Association of Universities for Research in Astronomy (AURA) under 
cooperative agreement with the National Science Foundation (NSF).}
software package. This includes  bias subtraction, 
flat--field correction, cosmic-ray removal, wavelength calibration, 
night sky background subtraction, correction for atmospheric extinction and 
absolute flux calibration of the two--dimensional spectrum.
For each night the sensitivity curve was derived from 
averaging of three standard stars. 

 One-dimensional spectra of the bright H {\sc ii} regions 
in each galaxy were corrected for interstellar extinction using the 
reddening curve by \citet{W58} and for
redshift, derived from the observed wavelengths
of the emission lines. Redshift-corrected spectra are shown in Fig.~\ref{sp_1}.

Emission line fluxes were measured using Gaussian profile fitting. 
The errors of the line flux measurements were calculated from the photon
statistics of non-flux-calibrated spectra. They have been propagated in the 
calculations of the elemental abundance errors.
 The observed relative emission line fluxes $F$($\lambda$)/$F$(H$\beta$)  
and fluxes $I$($\lambda$)/$I$(H$\beta$) corrected for interstellar 
extinction and underlying stellar absorption,  
equivalent widths EWs of emission lines, extinction coefficients 
$C$(H$\beta$), observed H$\beta$ fluxes $F$(H$\beta$), and equivalent widths of the hydrogen 
absorption lines are listed in Table \ref{t2_1}.
 Not all galaxies observed with the 3.6m telescope are included in 
Table \ref{t2_1}, 
but all of them are used in the subsequent analysis.
 We excluded from Table \ref{t2_1} 
eight H {\sc ii} regions 
with an oxygen abundance of 12 + log O/H $\leq$ 7.6. 
The emission line fluxes, electron temperatures, ionic and total element abundances 
for these H {\sc ii} regions are presented in Papaderos et al. (2006, in
preparation), as part of 
a detailed spectroscopic and photometric study of extremely metal-deficient star-forming galaxies.
  We neither include in Table \ref{t2_1} 
the two extremely 
metal-poor emission-line galaxies SDSS J2104--0035 and J0133+0052 discovered in SDSS. 
All observational data for these galaxies were published by \citet{IPGFT06a}.
  Finally, Table \ref{t2_1} 
does not include the extremely metal-deficient BCDs 
SBS 0335--052W and  SBS 0335--052E.
A spectroscopic study of these two systems is presented in  
\citet{PIGTF06}.

\section{Results \label{Results}}

\subsection{Electron temperature $T_e$({\rm O} {\sc iii}) and element 
abundances \label{T(OIII)}}


  \begin{figure}[t]
\hspace*{-0.0cm}\psfig{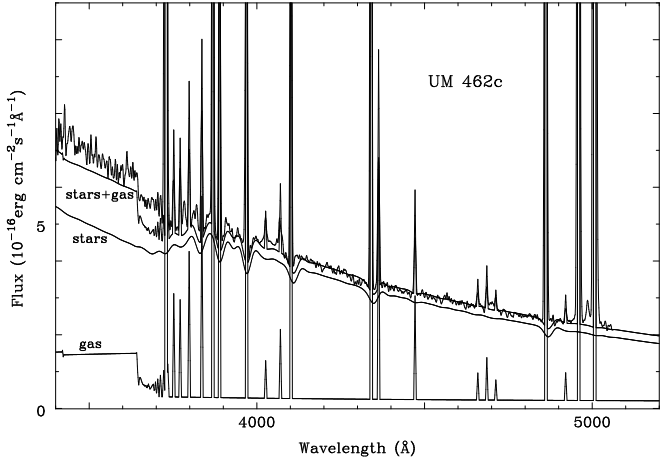}
    \caption{Best fit model SED to the redshift- and 
extinction-corrected observed spectrum of UM 462c. 
The model SED is calculated assuming that the 
H {\sc ii} region is ionisation-bounded.
 It is seen that the red part of spectrum is fitted quite well whereas 
the modelled SED underestimates the flux shortward of the Balmer jump.
  This is likely due to the leakage of ionising photons 
from the H {\sc ii} region.}
    \label{db_1}
\end{figure}

  \begin{figure*}[t]
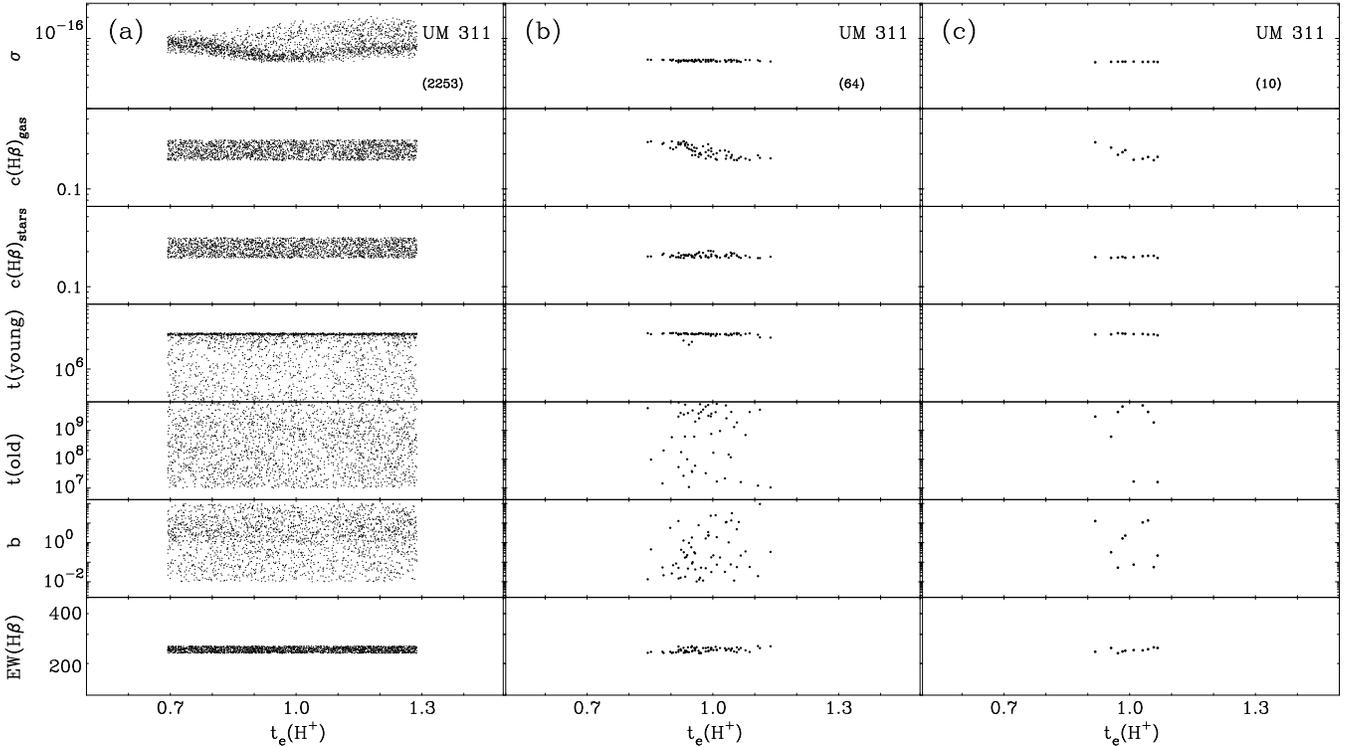


\hspace*{0.1cm}\psfig{figure=6067f2a.ps,angle=0,height=9.9cm,clip=}
\hspace*{-0.15cm}\psfig{figure=6067f2b.ps,angle=0,height=9.9cm,clip=}
\hspace*{-0.15cm}\psfig{figure=6067f2c.ps,angle=0,height=9.9cm,clip=}
    \caption{(a) Example of 
the parameter space explored with the Monte Carlo technique (2253 solutions 
out of 10$^5$ simulations) to fit the spectrum of the
BCD UM 311.
The parameters shown are the equivalent width of the H$\beta$ emission line EW(H$\beta$), 
the ages of the old and young stellar populations, $t$(old) and $t$(young),
the mass ratio $b$ of the old-to-young stellar population,
the extinction coefficient for the stellar emission $C$(H$\beta$)$_{stars}$,
the extinction coefficient for the ionised gas $C$(H$\beta$)$_{gas}$,
and the electron temperature $t_e$(H$^+$) = 10$^{-4}$$T_e$(H$^+$) in the H$^+$ zone.
The parameter $\sigma$ is an estimator of the goodness of the fit.
(b) and (c) are the same as (a)  except that only 
the 64 and 10 best Monte Carlo realizations are shown. 
}
    \label{db}
\end{figure*}

\setcounter{figure}{3}
  \begin{figure*}[hbtp]
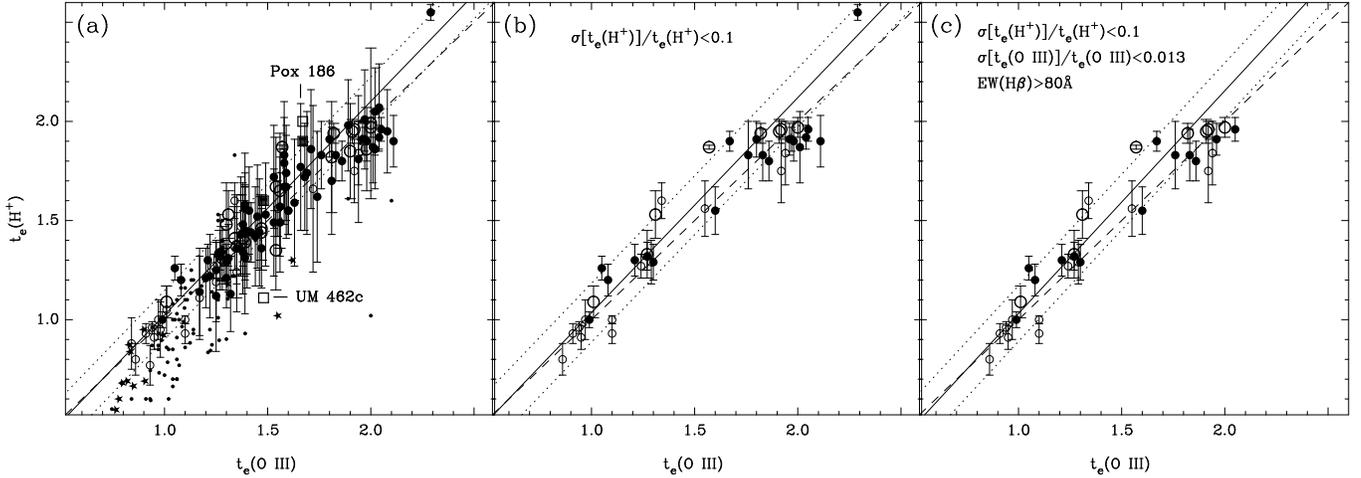

\hspace*{-0.05cm}\psfig{figure=6067f4a.ps,angle=0,height=6.3cm,clip=}
\hspace*{-0.15cm}\psfig{figure=6067f4b.ps,angle=0,height=6.3cm,clip=}
\hspace*{-0.15cm}\psfig{figure=6067f4c.ps,angle=0,height=6.3cm,clip=}
    \caption{
(a) Comparison of the $t_e$(H$^+$) derived from fitting the Balmer jump and
the SED with the $t_e$(O {\sc iii}) = 10$^{-4}$$T_e$(O {\sc iii}) 
derived from the nebular to auroral line 
flux ratio [O {\sc iii}] $\lambda$(4959+5007)/$\lambda$4363.
In each panel, the dashed line denotes equal temperatures,
the solid line is the linear regression obtained by the likelihood
method \citep{Pr92}, dotted lines denote 1$\sigma$ dispersions of our
sample H {\sc ii} regions around the regression line.
Error bars are the root mean square temperature deviations from the 
mean of the 10 best  Monte Carlo simulations. 
3.6m data are shown by large filled circles.
  Two objects, Pox 186 and UM 462c, which could not be fitted by 
ionisation-bounded model are denoted by squares. 
For each of these two galaxies we show by open squares the solutions 
from the ionisation-bounded model and by filled squares the solutions from 
the model with escaping Lyc photons. 
The derived $t_e$(H$^+$) from the Balmer jump in 23 
low-metallicity H {\sc ii} regions (MMT data) and the Paschen jump
in 24 low-metallicity H {\sc ii} regions (SDSS data)
are shown by large and small open circles, respectively \citep{GIT2006}.
   For comparison we have also plotted by dots in (a) 
the data of \citet{Liu2004}, \citet{Z2004}, \citet{W05} and \citet{KC05}
for 58 Galactic PNe.
Stars show the data for H {\sc ii} regions
from \citet{E98}, \citet{G04,G06}, \citet{P92}, \citet{P00}, \citet{P03} and 
\citet{G_D94}.
  (b) This panel is identical to panel (a) except that 
only H {\sc ii} regions with small dispersions for $t_e$(H$^+$)
($\sigma$[$t_e$(H$^+$)]$/$$t_e$(H$^+$) $<$ 0.1)
for 3.6m data (present paper) and data from \citet{GIT2006} (MMT and SDSS data) 
are shown. (c) The same as in (a) but only 
 H {\sc ii} regions with large equivalent widths and small errors 
for $t_e$(H$^+$) and $t_e$(O {\sc iii}) are included
($\sigma$[$t_e$(H$^+$)]$/$$t_e$(H$^+$) $<$ 0.1;
$\sigma$[$t_e$(O {\sc iii})]$/$$t_e$(O {\sc iii}) $<$ 0.013;
EW(H$\beta$) $>$ 80\AA).
 \label{t_result}}
\end{figure*}


The electron temperature $T_{\rm e}$, ionic and total heavy element abundances 
were derived following \citet{I06b}. 
  In particular, for the O$^{2+}$ and Ne$^{2+}$ ions we adopt
the temperature $T_e$(O {\sc iii}) derived from the 
[O {\sc iii}] $\lambda$4363/($\lambda$4959 + $\lambda$5007)
emission line ratio. 
  The  O$^+$ and  Fe$^{++}$ abundances were derived with the temperature
$T_e$(O {\sc ii}). 
 The latter was obtained from the relation between $T_e$(O {\sc iii})
and $T_e$(O {\sc ii}) of \citet{I06b}.
The 3.6m spectra covered only the blue wavelength region,
so that the [S {\sc ii}] $\lambda$6717, 6731 emission lines
can not be used for the electron number density determination.
  Therefore we have adopted  $N_e$ = 100 cm$^{-3}$.
The electron temperatures $T_{\rm e}$(O {\sc iii}) and $T_{\rm e}$(O {\sc ii}) 
for the high- and low-ionisation zones in H {\sc ii} regions respectively,
the ionisation correction factors ($ICF$s) and the 
ionic and total heavy element abundances for 
oxygen, neon and iron are given 
in Table~\ref{t3_1} excluding data for 16 H {\sc ii} regions 
published in the papers by \citet{IPGFT06a,PIGTF06} and Papaderos et al.
(2006, in preparation).

  Relevant parameters to the purpose of the paper 
such as the equivalent width EW(H$\beta$) of the H$\beta$ emission line, 
the electron temperature $t_{\rm e}$(O {\sc iii}) of the O$^{2+}$ zone, 
the electron temperature $t_{\rm e}$(H$^+$) of the H$^+$ zone,
the oxygen abundance 12 + log O/H and the extinction coefficient 
$C$(H$\beta$), obtained from the Balmer decrement, 
are collected in Table~\ref{te} for the entire sample 
of the 69 H {\sc ii} regions observed with the 3.6m ESO telescope.

\subsection{Electron temperature $T_e$({\rm H}$^+$)  \label{method}}

  To derive the electron temperature $T_e$(H$^+$) we use the same method 
as that described in detail  by \citet{GIT2006}.
  The method is based on the determination of the electron temperature $T_e$(H$^+$) 
of H$^+$ zones
by fitting a series of model SEDs to the 
observed SEDs and finding the best fits. 
 Each fit is performed 
over the whole observed spectral range shortward of
$\sim$ $\lambda$(5100 -- 5200)\AA, which includes the Balmer jump region ($\lambda$3646\AA).
Besides the electron temperature $T_e$(H$^+$) which controls the magnitude 
of the Balmer jump, the shape of the SED depends on several other parameters.
As each SED is the sum of both stellar and ionised gas emission, its shape depends 
on the relative brightness of these two components. 
  In BCDs, the contribution of the ionised gas emission can be very large.
 However, the EWs of hydrogen emission lines never attain the theoretical values for pure 
ionized gas emission. This implies a non-negligible contribution of stellar emission in all
sample galaxies. 
We therefore parametrize the relative  contribution of gaseous emission to the stellar 
one by the equivalent width EW(H$\beta$).

The shape of the spectrum depends also on reddening.
The extinction coefficient for the ionised gas $C$(H$\beta$)
has been obtained from the observed hydrogen Balmer decrement. 
Here we assume that the differences between the observed and theoretical
recombination hydrogen Balmer line ratios are only due to reddening and 
underlying stellar absorption. However, additional source of differences,
collisional excitation of hydrogen lines, could play some role in the hot
H {\sc ii} regions \citep{SI03,Lur03b}. Collisional excitation may increase 
the flux of the H$\alpha$ emission line by $\la$ 5\% above the recombination
value. The effect is less important, $\la$ 2\%, for the H$\beta$ emission
line. 
 We find that ignoring collisional excitation of hydrogen lines does not
change significantly the derived $T_e$(H$^+$). As the H$\alpha$ line is not 
present in our spectra, we decreased by 2\% the flux of H$\beta$ line measured
in spectra of several hottest H {\sc ii} regions from our sample. 
Such correction results in decreasing $C$(H$\beta$) by $\sim$ 0.07.
  Then the derived $T_e$(H$^+$) is not changed more than by $\sim$ 1\% as 
compared to the case with the measured flux of the H$\beta$ line.
  Therefore, for the whole sample we decided not to take into account collisional
excitation of hydrogen lines and use the $C$(H$\beta$) obtained from the
measured hydrogen line fluxes.
 We have no direct observational constraint for the reddening 
for the stellar component, which could differ from $C$(H$\beta$).
Therefore, for simplicity, we adopt in calculations that both extinctions
$C$(H$\beta$)$_{gas}$ and $C$(H$\beta$)$_{stars}$, respectively for the ionised
gas emission and the stellar component emission, are randomly varied in the 
narrow range around $C$(H$\beta$).
   Finally, the SED depends on the star formation history of the BCD.

We have carried out a series of Monte Carlo simulations to reproduce the SED
in each H {\sc ii} region of our sample. 
   To calculate the contribution of stellar emission to the SEDs,
we have adopted a grid of the Padua stellar evolution
models by 
\citet{Gi00}\footnote{http://pleiadi.pd.astro.it.} 
with heavy
element mass fractions $Z$ = 0.0001, 0.0004, 0.001, 0.004, 0.008. 
  Using these data we have calculated with the
package PEGASE.2 \citep{FR97} a grid of  
instantaneous burst SEDs in a wide range of ages, from  0.5 Myr to 15 Gyr.
 We have adopted a stellar initial mass function with a Salpeter 
slope, an
upper mass limit of 100 $M_\odot$ and a lower mass limit of 0.1 $M_\odot$.
  Following \citet{GIT2006} we could approximate the  
star formation history in BCDs   
by two short bursts of different age but equal durations
and with different strengths.
   The old stellar population is usually approximated by continuous star formation 
\citep[e.g. ][]{Guseva2001,Guseva2003a,Guseva2003b,Guseva2003c}.
 However, the contribution of the old stellar population to the light
of the bright H {\sc ii} regions considered in this paper is always small, not exceeding a few
percent in the optical range. 
   Therefore, the approximation of a short burst for the old stellar population
is acceptable for our modelling, as it was shown by \citet{GIT2006}.
   The contribution of gaseous emission to the total emission is scaled by the ratio
of the observed equivalent width of the H$\beta$ emission line
to the EW(H$\beta$) expected for pure gaseous emission.
  The gaseous continuum emission is calculated following \citet{Aller84}
and includes hydrogen and helium free-bound, free-free, and two-photon emission.
In our models it is always calculated with the electron temperature $T_e$(H$^+$) 
and with the He/H abundance ratio derived from the H {\sc ii} region spectrum.
The observed emission lines corrected for reddening and scaled using the absolute flux of the
H$\beta$ emission line were added to the calculated gaseous continuum.

Given that $T_e$(H$^+$) is not necessarily equal to $T_e$(O {\sc iii}), 
we chose to vary it in the range (0.7 -- 1.3)$\times$$T_e$(O {\sc iii}). 
We adopt for the range of the allowed EW(H$\beta$) values between 0.95 and 1.05 
times its nominal value.
  As for the extinction, we allow both  $C$(H$\beta$)$_{stars}$ and $C$(H$\beta$)$_{gas}$ 
to vary in the range of (0.8 -- 1.2) of $C$(H$\beta$), obtained from the Balmer decrement.

  We assume that the young stellar population was formed in a recent burst of 
star formation with an age $t$(young) between 0.3 and 10 Myr.  
For the age of the older stars $t$(old) we adopt values between 10 Myr and 15 Gyr.
The contribution of each burst to the SED is defined by the ratio of the masses of stellar populations
formed respectively in the old and young bursts, $b$ = $M$(old)/$M$(young), which we vary 
between 0.01 and 100. 
  For each H {\sc ii} region we computed a series of synthetic SEDs
using a grid of Padua models with the heavy element mass fraction,
that is closest to the heavy element mass fraction of the ionised gas.
We run 10$^5$  Monte Carlo models for each H {\sc ii} region 
varying simultaneously $t$(young), $t$(old), $b$, $T_e$(H$^+$),
$C$(H$\beta$)$_{gas}$ and $C$(H$\beta$)$_{stars}$.
 The first four parameters are used to 
calculate the model EW(H$\beta$).
Only those solutions were further considered for the SED fitting in which 
the modelled EW(H$\beta$) falls in the range 0.95 -- 1.05 of 
the observed one. 
  Typically, the number of such solutions is $\sim$ 10 -- 100 times less 
than the total of 10$^5$ simulations.

In the same way as in \citet{GIT2006} we used the $\sigma$ statistic
to quantify the goodness of each model's fit to the observed SED.
 For each Monte Carlo realization we computed a mean deviation $\sigma$ 
between the observed and modelled spectra.
  We calculate in each spectrum the root mean square deviation $\sigma_j$ for  
each of the five continuum regions selected to be devoid of line emission
and absorption, 
as $\sigma_j$ = $ \sqrt{\sum_{i=1}^{N_j}{(f^i_{obs}-f^i_{mod})^2} / N_j}$, where $N_j$ is 
the number of the points in each particular spectral interval. 
Then $\sigma$ = $\sum{\sigma_j}/5$.

  As shown in Fig. 2a, for all Monte Carlo realizations the age 
$t$(old) and $t$(young) and mass ratio $b$ of the old and young 
stellar populations, and the electron temperature $T_e$(H$^+$) in 
the H$^+$ zone span a wide range.
However, if only 
solutions with the lowest $\sigma$s are taken into account then the range of 
the electron temperature in the H$^+$ zone for many H {\sc ii} regions
from our sample is considerably narrowed (Fig.~\ref{db}b and ~\ref{db}c). 
 On the other hand, for some H {\sc ii} regions the minimum in the 
distribution of $\sigma$ is
not sharp and even two minima in the distribution of $\sigma$ are obtained 
with significantly different $T_e$(H$^+$) 
\citep[see also Fig. 9 in ][]{GIT2006}. 
The $T_e$(H$^+$) of each H {\sc ii} region is derived as 
the average of the electron temperatures obtained 
for several best-fitting SEDs. We also derive the dispersion 
around the average $T_e$(H$^+$) of the electron 
temperatures obtained from the same best-fitting SEDs. Then small dispersions 
correspond to sharp single minimuma in the distributions of $\sigma$. 
  We have checked how the derived $T_e$(H$^+$) does depend 
on the exact number of best solutions we choose to average.
No systematic differences were found when the number 
of best fits varies between a few and a couple of ten \citep{GIT2006}.
  Hence, we adopt for $t_e$(H$^+$) = 10$^{-4}$$T_e$(H$^+$) the 
mean of the 10 best-fitting solutions weighted by $\sigma$.

We note that the bimodal $\sigma$ distribution for some H {\sc ii} 
regions does not allow to derive a unique $T_e$(H$^+$) in those objects. 
The average $T_e$(H$^+$) of the 10 best-fitting solutions for these H {\sc ii} 
regions can significantly differ from the actual one. On the other hand, 
the dispersion of the average $T_e$(H$^+$)s in H {\sc ii} regions with the 
bimodal $\sigma$ distribution is large reflecting
the uncertainties of the temperature determination. Therefore,
in the following discussion we will use different samples: the total sample 
(Fig. \ref{t_result}a) and 
the samples with the sharp minimum in the $\sigma$ distribution
(Figs. \ref{t_result}b,c).

\subsection{Comparison of $t_e$(H$^+$) and $t_e$(O {\sc iii})}

  Figure~\ref{sp_1} shows by a thick solid line the best-fit SED
(SED with the smallest $\sigma$) 
superimposed on the redshift- and extinction-corrected observed spectrum of 
each H {\sc ii} region in the sample, 
including the H {\sc ii} regions discussed by \citet{IPGFT06a} and 
Papaderos et al. (2006, in preparation).

 The galaxies are arranged in order of decreasing H$\beta$ equivalent 
width, i.e. in order of decreasing contribution of gaseous emission 
relative to stellar emission.
  It is seen that, in every case, the model SED fits well the observed one over 
the whole spectral range, including the Balmer jump region.  
The separate contributions 
from the stellar and ionised gas components are shown 
only for SBS 0335--052E SE 
by thin solid lines.

In Fig.~\ref{t_result}a we compare $t_e$(H$^+$)
and $t_e$(O {\sc iii}) = 10$^{-4}$ $T_e$(O {\sc iii}).
  The 3.6m data for the sample of 69 H {\sc ii} regions in 45 BCDs
are shown by large filled circles and, in the case of Pox 186 and UM 462c, by squares.
  The electron temperature $t_e$(H$^+$) for each galaxy is the  
mean of the temperatures derived from the 10 best Monte Carlo 
realizations weighted by their $\sigma$s (Table~\ref{te}).
Error bars are the root mean square temperature deviations from the 
mean of the 10 best  Monte Carlo simulations. 
The dashed line denotes equal temperatures, while
the solid line is a linear regression 
$t_e$(H$^+$) = 1.078$\times$$t_e$(O {\sc iii}) -- 0.054
obtained with the likelihood method \citep{Pr92}
for all 116 H {\sc ii} regions
that takes into account errors in $t_e$(H$^+$) and $t_e$(O {\sc iii})
in each object. 
  Dotted lines illustrate 1$\sigma$ deviations 
of the sample H {\sc ii} regions from the linear regression. The linear
regression is almost not changed if the outlying object with the highest
electron temperature $t_e$(O {\sc iii}) is excluded. 
It is seen from
Fig.~\ref{t_result}a that the linear regression (solid line) is slightly 
different from the line of equal temperatures (dashed line) because of the
small dispersion of $t_e$(H$^+$) for the H {\sc ii} region Mrk 71 No.1
\citep{GIT2006} and, therefore, its large weight in the linear regression
determination. On the other hand, if equal dispersions of $t_e$(H$^+$)
are adopted for all H {\sc ii} regions, the linear regression deviates
very little from the line of equal temperatures. All these 
differences are smaller than the 1$\sigma$ deviations of H {\sc ii} regions
from the linear regression (dotted lines). Therefore, we conclude that
no systematic difference between $t_e$(O {\sc iii}) and $t_e$(H$^+$) is 
evident from  Fig.~\ref{t_result}a.

  Our 3.6m observations in Fig.~\ref{t_result}a are supplemented by the data
from previous studies.
 The results  of $t_e$(H$^+$) determination from Balmer jump for 
23 low-metallicity H {\sc ii} regions (MMT data) and Paschen jump
for 24 low-metallicity H {\sc ii} regions (SDSS data) from \citet{GIT2006}
are shown by large and small open circles respectively.
 In Fig.~\ref{t_result}a we also plot some other data from
the literature.
  By stars are shown the  
H {\sc ii} regions:  M17 \citep{P92}, Orion nebula
\citep{E98}, NGC 3576 \citep{G04}, S 310 \citep{G04,G06} in the Galaxy;  
30 Dor in Large Magellanic Cloud \citep{P03};  NGC 346 
in Small Magellanic Cloud \citep{P00}; and H {\sc ii} regions 
A and B in the BCD Mrk 71 \citep{G_D94}. 
   By dots in Fig. ~\ref{t_result}a we plot the data  
for 58 Galactic PNe by \citet{Liu2004}, 
\citet{Z2004},  \citet{W05} and \citet{KC05}.
 The data collected from literature 
extend the $t_e$(H$^+$) -- $t_e$(O {\sc iii})
relation to lower temperatures. 

  In  Fig.~\ref{t_result}b are shown only 41 H {\sc ii} regions 
with small dispersions of $t_e$(H$^+$)
($\sigma$[$t_e$(H$^+$)]$/$$t_e$(H$^+$) $<$ 0.1)
for 3.6m data (present paper) and for data from \citet{GIT2006} (MMT and SDSS data). 
These data are fitted by a linear regression 
$t_e$(H$^+$) = 1.085$\times$$t_e$(O {\sc iii}) -- 0.050 (solid line) that is
very similar to the linear regression obtained for the whole sample of
116 H {\sc ii} region. If the outlying point with the highest 
$t_e$(O {\sc iii}) is excluded the linear regression is almost not changed.
Thus, from the analysis of the subsample of H {\sc ii} regions with
sharpest minima in the $\sigma$ distribution we reach the same conclusion
as that based on the total sample: there is no significant differences
between $t_e$(H$^+$) and $t_e$(O {\sc iii}). However, we cannot exclude small
differences of $\leq$ 1000 K, corresponding to the temperature fluctuation
parameter $t^2$ $\leq$ 0.02.

  Special emphasis was given to the analysis of a few H {\sc ii} 
regions which are most deviant from the line of equal electron temperatures 
in Fig.~\ref{t_result}a. Two of these H {\sc ii} regions, UM 462c and Pox 186, 
are labeled in Fig.~\ref{t_result}a. 
 One of these sources, Pox 186, has been studied previously by \citet{GPIet2004}.
No H {\sc i} 21 cm emission was detected in this galaxy. 
Its H {\sc ii} region is characterised by a very high O$^{2+}$/O$^+$ abundance 
ratio of $\sim$20 and likely shows Ly$\alpha$ emission in the UV. 
These properties led \citet{GPIet2004} to conclude that the H {\sc ii} region in Pox 186 
may be density-bounded.
If this is indeed the case then EW(H$\beta$) does not 
measure the starburst age anymore. Its inclusion into our modelling procedure
would then result in an incorrect reproduction of the star formation history and 
the stellar SED.
  While this assumption is less critical in the case of Pox 186 because 
of the high EW(H$\beta$) and hence the dominant contribution of gaseous emission 
to the total emission of this system, it is important for the case 
of a H {\sc ii} region with lower EW(H$\beta$), such as UM 462c.
    This is illustrated in Fig.~\ref{db_1} where we show the model SED which best fits the 
observed spectrum of UM 462c.
It is seen that the ionisation-bounded model of the H {\sc ii} region fails to reproduce the
observations regardless of the adopted synthetic SED.
Specifically, the models underestimate the SED blueward of the Balmer jump, 
whereas redward of it they provide a satisfactorily fit.
The most likely reason for this discrepancy is leakage of ionising photons 
from the H {\sc ii} region.

The discrepancy between observed and modelled SEDs can be 
eliminated, however, assuming a 
H {\sc ii} region model with escaping Lyc photons, which implies a larger number of 
massive ionising stars than the number needed to account for the observed 
EW(H$\beta$) in the case of an ionisation-bounded H {\sc ii} region.
  Thus, Monte Carlo simulations assuming a H {\sc ii} region 
with escaping Lyc photons require
a younger stellar population with a larger production rate of ionising photons, 
compared to models assuming an ionisation-bounded H {\sc ii} region.
  As a result the stellar SED blueward of the Balmer jump 
is actually larger than what fits to EW(H$\beta$) imply.    
  In practice, in order to account for the fraction of photons escaping 
from the H {\sc ii} region we need to introduce an additional parameter $f$.
Then, the modelled SED of the ionised gas emission 
and EW(H$\beta$) must be multiplied by a factor
(1 -- $f$) to match the observed ones.

  In Fig.~\ref{t_result}a we show by open squares the $t_e$(H$^+$) 
for Pox 186 and UM 462c obtained from the ionisation-bounded  model
and by filled squares the $t_e$(H$^+$) obtained from the 
model with escaping Lyc photons. 
It is seen that the $t_e$(H$^+$) obtained from 
the latter models for both H {\sc ii} regions (filled squares)
are in better agreement with the data for other H {\sc ii} regions.
This is also evident from  Fig.~\ref{sp_1} where we plot the 
best-fit SEDs obtained for these two systems from models
with escaping Lyc photons.
  These SEDs reproduce quite well the observed spectra over their 
whole wavelength range, including the region blueward of the Balmer jump.

The leakage of Lyc photons could be present in other
H {\sc ii} regions from our sample.
  This may introduce an additional source of uncertainties
in the determination of $t_e$(H$^+$). However, the fact that majority of 
H {\sc ii} regions are well fitted by the models with non-escaping Lyc photons 
suggests that in general $f$ is small. 
Our conclusion is supported by multi-wave-band studies by  \citet{L95}.
They find that less than 3\% of ionizing photons escape from local starburst 
galaxies.
 More recently, for several local starburst galaxies, \citet{Heck2001} 
estimate $f$ $\la$ 6\%
and \citet{Berg2006} find $f$ $\sim$ 4\% -- 10\% for another local starburst.
The direct detection of Lyman continuum emission from 14 high-redshift star-forming 
galaxies shows that only 2 of 14 studied galaxies have a significant emission 
below the Lyman limit \citep{Shap2006}.
 Therefore, for all our H {\sc ii} regions except for Pox 186 and UM 462c
we decided to apply the models with the smallest number of free parameters 
adopting $f$ = 0.

  Inspection of Fig.~\ref{t_result}b also reveals that some of the H {\sc ii} regions from the
3.6m data sample with the highest temperatures [($t_e$(H$^+$) $\sim$ 1.9 and 
$t_e$(O {\sc iii}) $\ga$ 2.0)] are offset from the 
linear regression and from the 
line of equal temperatures, at variance to
H {\sc ii} regions with a lower $t_e$(H$^+$). These offset galaxies are low-excitation 
H {\sc ii} regions with weak [O {\sc iii}]$\lambda$4363 emission lines.
Therefore, the determination of $t_e$(O {\sc iii}) is more uncertain in those galaxies.
  Indeed, the number of deviant points from the line of 
equal temperatures are decreased
if only galaxies with high equivalent widths 
of H$\beta$ (EW(H$\beta$) $>$ 80\AA) and small  
$t_e$(O {\sc iii}) and $t_e$(H$^+$) errors are considered (Fig.~\ref{t_result}c).

With the additional PNe data an interesting trend appears in Fig.~\ref{t_result}a. 
While the PNe data scatters nicely on either side of 
the line of equal temperatures for $t_e$(O {\sc iii}) $\ga$ 1.1, just as 
the BCD data, the  $t_e$(H$^+$) -- $t_e$(O {\sc iii}) relationship 
curves down for objects with $t_e$(O {\sc iii}) $\la$ 1.1, with 
$t_e$(H$^+$) being systematically lower than $t_e$(O {\sc iii}). 
 There are  several cool  H {\sc ii} regions 
collected from literature and shown by stars, 
which also follow the trend delineated by PNe. 
  All other H {\sc ii} regions with  $t_e$ $>$ 1.0 collected from literature 
do not deviate from line of equal temperatures,
similar to the sample of 116 H {\sc ii} regions
in 88 emission-line galaxies collected from  3.6m, MMT and SDSS data.

Thus, we confirm the previous  finding by \citet{GIT2006} that the
temperature of the O {\sc iii} zone is equal, within the errors, 
to the temperature of the H$^+$ zone in the H {\sc ii} regions with
$T_e$(H$^+$) $\ga$ 11000 K.
  We emphasize that, due to the large observational uncertainties for each 
individual object and a number of modelling assumptions,
only a statistical study of a large, homogeneous sample of H {\sc ii} 
regions allows us to firmly establish the relation between $T_e$(H$^+$) and
$T_e$(O {\sc iii}).

 \section{Conclusions \label{conc}}

Based on 3.6m ESO spectroscopic data, we have determined 
in this paper the temperatures of the H$^+$ zones  
in 69 H {\sc ii} regions of 45 blue compact dwarf (BCD) galaxies.
  Following the procedure of \citet{GIT2006}, we have used a Monte Carlo 
technique to vary free parameters and
to calculate a series of model spectral energy distributions (SEDs). 
The electron temperature of H$^+$ zones were derived from best-fit SEDs to the 
observed spectra in the wavelength range $\sim$ 3200--5100\AA, which includes 
the  Balmer jump region. 
   These new data are combined with a sample of 47  H {\sc ii} regions from  \citet{GIT2006}
in order to study  systematic trends in the electron temperature determination 
over a wide range in oxygen abundance (7.13 $\leq$ 12+log(O/H) $\leq$ 8.36) 
and H$\beta$ equivalent width (10\AA\ $\leq$EW(H$\beta$) $\leq$397\AA).
This is the largest spectroscopic sample of low-metallicity H {\sc ii} regions 
ever compiled in order to address the long-standing 
question of how the electron temperature of H$^+$ zones is related to that 
derived from collisionally excited forbidden lines.

We conclude that in H {\sc ii} regions with an 
electron temperature $T_e$(H$^+$) $\ga$ 11000 K the
temperature of the O$^{2+}$ zone, determined from the nebular to auroral line 
flux ratio of doubly ionised oxygen [O {\sc iii}] $\lambda$(4959+5007)/$\lambda$4363 is, 
within the errors, equal to the temperature of the H$^+$ zone determined from 
fitting the SED including that in the Balmer jump wavelenght region. 
However, we could not exclude small differences of $\leq$ 1000 K between
$T_e$(H$^+$) and $T_e$(O {\sc iii}) corresponding to the temperature
fluctuations parameter $t^2$ $\leq$ 0.02.

Consequently, temperature differences could not be of concern for the 
determination of the primordial He abundance, as it relies mainly on the 
hot low-metallicity H {\sc ii} regions. 
  Thus, on the base of a new large sample of low-metallicity BCDs  
we confirm and strengthen the previous finding by \citet{GIT2006}, 
based on the analysis and modelling of the Balmer and Paschen jump in a smaller sample
of low-metallicity H {\sc ii} regions.

 We also identify and discuss cases of H {\sc ii} regions where ionisation-bounded models 
fail to fit the observed SEDs.
Such models typically underestimate the object's SED blueward of the Balmer discontinuity.
This can be, however, plausibly accounted for by photoionisation models allowing 
for the partial leakage of Lyc photons from a H {\sc ii} region.

\acknowledgements
Y. I. I. and N. G. G. thank the hospitality of the Institute for Astrophysics
(G\"ottingen), and the support of the DFG grant No. 436 UKR 17/25/05.
P.P. would like to thank Gaspare Lo Curto, Lorenzo Monaco, Carlos La Fuente, 
Eduardo Matamoros and the whole ESO staff at the La Silla Observatory for their support.
All the authors acknowledge the work of the Sloan Digital Sky
Survey (SDSS) team.
Funding for the SDSS has been provided by the
Alfred P. Sloan Foundation, the Participating Institutions, the National
Aeronautics and Space Administration, the National Science Foundation, the
U.S. Department of Energy, the Japanese Monbukagakusho, and the Max Planck
Society. The SDSS Web site is http://www.sdss.org/.


\setcounter{table}{1}


\begin{longtable}{llcccc} 
\caption{Comparison of $t_e$(O {\sc iii})$^a$ 
and $t_e$(H$^+$)$^b$ for the sample galaxies \label{te}}\\
\hline\hline
\endfirsthead
\caption{continued.}\\
\hline\hline
Galaxy          & EW(H$\beta$) & $t_e$(O {\sc iii}) 
& $t_e$(H$^+$) & 12 + log O/H 
& $C$(H$\beta$)$^c$ \\ \hline
\endhead
\hline
\endfoot
Galaxy          & EW(H$\beta$) & $t_e$(O {\sc iii}) 
& $t_e$(H$^+$) & 12 + log O/H 
& $C$(H$\beta$)$^c$ \\ 
\hline
SBS 0335-052E SE  & 382. & 1.98$\pm$0.05 & 1.90$\pm$0.10&  7.28 &  0.00   \\  
Pox 186           & 358. & 1.67$\pm$0.02 & 2.00$\pm$0.07&  7.75 &  0.08   \\  
HS 2236+1344a     & 343. & 2.05$\pm$0.02 & 1.96$\pm$0.06&  7.49 &  0.14   \\  
UM 461a           & 335. & 1.66$\pm$0.02 & 1.77$\pm$0.32&  7.76 &  0.22   \\  
J 104457.80+035313.1a      & 318. & 1.96$\pm$0.02 & 1.91$\pm$0.08&  7.44 &  0.31   \\  
Tol 1214-277      & 314. & 2.04$\pm$0.03 & 1.92$\pm$0.06&  7.50 &  0.14   \\  
UM 570            & 310. & 1.83$\pm$0.02 & 1.83$\pm$0.13&  7.72 &  0.14   \\  
CGCG 032-017       & 308. & 1.38$\pm$0.01 & 1.48$\pm$0.21&  8.09 &  0.27   \\ 
Pox 108           & 306. & 1.31$\pm$0.02 & 1.31$\pm$0.17&  8.11 &  0.08   \\  
UM 559            & 306. & 1.60$\pm$0.02 & 1.55$\pm$0.12&  7.72 &  0.06   \\    
SBS 0335-052 \#7  & 299. & 1.97$\pm$0.06 & 1.91$\pm$0.06&  7.21 &  0.14    \\ 
UM 442            & 289. & 1.30$\pm$0.01 & 1.29$\pm$0.09&  8.10 &  0.14    \\ 
Mrk 1329          & 288. & 1.05$\pm$0.01 & 1.26$\pm$0.06&  8.33 &  0.03   \\  
Mrk 1236a         & 283. & 1.21$\pm$0.01 & 1.30$\pm$0.08&  8.17 &  0.14   \\  
Tol 2146-391      & 277. & 1.63$\pm$0.02 & 1.59$\pm$0.32&  7.78 &  0.18   \\ 
CGCG 007-025a      & 274. & 1.58$\pm$0.02 & 1.83$\pm$0.27&  7.81 &  0.20   \\  
HS 2134+0400      & 262. & 2.02$\pm$0.04 & 1.86$\pm$0.20&  7.41 &  0.25    \\ 
J 001434.98-004352.0       & 260. & 1.39$\pm$0.04 & 1.31$\pm$0.28&  7.97 &  0.12    \\ 
J 232420.34-000625.0       & 249. & 1.41$\pm$0.01 & 1.55$\pm$0.22&  7.95 &  0.20   \\  
SBS 0335-052E \#3b    & 249. & 2.04$\pm$0.03 & 2.07$\pm$0.22&  7.31 &  0.45   \\  
J 125305.96-031258.9       & 248. & 1.38$\pm$0.01 & 1.34$\pm$0.21&  8.01 &  0.36   \\  
J 120122.30+021108.3       & 248. & 1.81$\pm$0.03 & 1.70$\pm$0.27&  7.49 &  0.09   \\  
SBS 0335-052E \#3a    & 246. & 2.02$\pm$0.03 & 2.05$\pm$0.22&  7.31 &  0.43   \\  
Tol 65a           & 244. & 1.76$\pm$0.02 & 1.83$\pm$0.17&  7.54 &  0.10   \\  
UM311             & 243. & 0.99$\pm$0.01 & 1.00$\pm$0.04&  8.36 &  0.22   \\  
J 130432.27-033322.1       & 233. & 1.08$\pm$0.01 & 1.20$\pm$0.08&  8.24 &  0.12   \\  
J 024815.95-081716.5a      & 233. & 1.32$\pm$0.01 & 1.13$\pm$0.19&  8.04 &  0.04   \\  
Pox 4a            & 220. & 1.39$\pm$0.01 & 1.11$\pm$0.23&  8.01 &  0.08   \\  
Pox 120a          & 215. & 1.53$\pm$0.01 & 1.76$\pm$0.22&  7.88 &  0.21   \\  
HS 2236+1344b     & 215. & 1.86$\pm$0.02 & 1.80$\pm$0.10&  7.58 &  0.22    \\ 
CGCG 007-025c      & 214. & 1.56$\pm$0.02 & 1.49$\pm$0.25&  7.78 &  0.02    \\ 
UM 462a           & 212. & 1.39$\pm$0.01 & 1.45$\pm$0.22&  8.00 &  0.52   \\  
J 230210.00+004938.8       & 212. & 1.80$\pm$0.03 & 1.91$\pm$0.09&  7.63 &  0.00   \\  
Tol 2138-405      & 208. & 1.39$\pm$0.02 & 1.08$\pm$0.12&  7.99 &  0.39   \\  
J 210455.31-003522.2N       & 203. & 2.01$\pm$0.08 & 1.87$\pm$0.18&  7.26 &  0.00   \\  
J 011340.44+005239.1       & 193. & 2.29$\pm$0.30 & 2.55$\pm$0.04&  7.17 &  0.00   \\  
J 013352.56+134209.3       & 178. & 1.74$\pm$0.02 & 1.62$\pm$0.33&  7.56 &  0.04   \\  
Pox 4b            & 177. & 1.43$\pm$0.01 & 1.51$\pm$0.18&  7.87 &  0.00   \\  
IC 4662a          & 172. & 1.27$\pm$0.01 & 1.34$\pm$0.19&  8.09 &  0.44   \\  
J 143053.50+002746.3       & 168. & 1.30$\pm$0.01 & 1.21$\pm$0.15&  8.07 &  0.15   \\  
IC 4662c          & 166. & 1.25$\pm$0.01 & 1.25$\pm$0.14&  8.16 &  0.33   \\  
Pox 120b          & 158. & 1.37$\pm$0.02 & 1.18$\pm$0.20&  7.99 &  0.14   \\  
g2252292-171050   & 158. & 1.59$\pm$0.02 & 1.74$\pm$0.24&  7.82 &  0.10    \\ 
UM 461b           & 147. & 1.58$\pm$0.02 & 1.79$\pm$0.23&  7.77 &  0.18   \\  
Mrk 1236c         & 147. & 1.22$\pm$0.04 & 0.98$\pm$0.03&  8.05 &  0.00   \\  
J 144805.36-011057.7       & 146. & 1.39$\pm$0.01 & 1.44$\pm$0.24&  8.00 &  0.21   \\  
IC 4662b          & 140. & 1.27$\pm$0.01 & 1.32$\pm$0.07&  8.10 &  0.14   \\  
J 223036.79-000636.9       & 133. & 1.71$\pm$0.03 & 1.86$\pm$0.30&  7.64 &  0.34   \\  
Pox 105           & 131. & 1.53$\pm$0.02 & 1.49$\pm$0.27&  7.85 &  0.04   \\  
UM 462d           & 130. & 1.45$\pm$0.03 & 1.52$\pm$0.26&  7.86 &  0.23   \\  
UM 462c           & 127. & 1.48$\pm$0.02 & 1.11$\pm$0.15&  7.88 &  0.00   \\  
J 024815.95-081716.5b      & 125. & 1.35$\pm$0.04 & 1.36$\pm$0.18&  7.99 &  0.00   \\  
CGCG 007-025b      & 122. & 1.58$\pm$0.02 & 1.67$\pm$0.19&  7.75 &  0.06    \\ 
J 162410.12-002202.5       & 122. & 1.20$\pm$0.01 & 1.21$\pm$0.15&  8.16 &  0.16   \\ 
UM 462e           & 112. & 1.59$\pm$0.05 & 1.67$\pm$0.26&  7.77 &  0.00   \\  
J 013452.00-003854.4       & 111. & 1.49$\pm$0.15 & 1.53$\pm$0.26&  7.85 &  0.16   \\  
SBS 0335-052W     & 109. & 1.97$\pm$0.25 & 2.01$\pm$0.25&  7.13 &  0.02   \\  
Tol 65b           &  96. & 1.69$\pm$0.04 & 1.74$\pm$0.31&  7.59 &  0.14   \\  
SBS 0335-052E \#4,5    &  89. & 2.11$\pm$0.04 & 1.90$\pm$0.13&  7.27 &  0.26   \\  
UM 462b           &  85. & 1.42$\pm$0.02 & 1.44$\pm$0.15&  7.92 &  0.21   \\  
CGCG 007-025d      &  69. & 1.68$\pm$0.04 & 1.72$\pm$0.27&  7.65 &  0.00   \\  
ESO 338-IG04      &  59. & 1.46$\pm$0.02 & 1.44$\pm$0.27&  7.90 &  0.02   \\  
Mrk 1236b         &  56. & 1.17$\pm$0.02 & 1.14$\pm$0.22&  8.12 &  0.07    \\ 
J 141454.13-020822.9       &  48. & 1.89$\pm$0.01 & 1.98$\pm$0.23&  7.32 &  0.08    \\ 
IC 4662d          &  44. & 1.25$\pm$0.02 & 1.12$\pm$0.28&  8.07 &  0.14    \\ 
J 010746.56+010352.0       &  40. & 1.56$\pm$0.23 & 1.57$\pm$0.25&  7.68 &  0.12    \\ 
g0405204-364859   &  23. & 2.08$\pm$0.25 & 1.95$\pm$0.21&  7.34 &  0.02    \\ 
J 104457.80+035313.1b      &  20. & 1.94$\pm$0.21 & 1.81$\pm$0.32&  7.45 &  0.14    \\ 
IC 4662e          &  10. & 1.31$\pm$0.01 & 1.30$\pm$0.23&  8.08 &  0.06    \\ 
\hline
\multicolumn{6}{l}{$^a$$t_e$(O {\sc iii}) = 
$T_e$(O {\sc iii})/10000 K, derived from the [O {\sc iii}] $\lambda$(4959+5007)/$\lambda$4363 emission line
ratio.} \\
\multicolumn{6}{l}{$^b$$t_e$(H$^+$) = $T_e$(H$^+$)/10000, derived as the 
weighted mean temperature from the 10 best fits of the SED.} \\
\multicolumn{6}{l}{$^c$Extinction coefficient is derived from observed hydrogen 
Balmer decrement.}
\end{longtable}



\Online

\setcounter{table}{0}


\begin{table*}  
\caption{Journal of the observations
\label{journal}}
\begin{tabular}{llcc|llcc} \hline  \hline
Name$^a$& Data of Obs & Exp$^b$ (s) & A.M.$^c$& Name$^a$ &  Data of Obs & Exp$^b$ (s) & A.M.$^c$ \\ \hline
SBS 0335-052E SE     & Oct 9   & 2700 & 1.16  & J 0113+0052          & Oct 9  & 3600 & 1.21  \\
Pox 186              & Apr 14  & 2400 & 1.27  & J 0133+1342          & Oct 7  & 3600 & 1.38  \\
HS 2236+1344a        & Oct 7   & 3600 & 1.38  & Pox 4b               & Apr 12 & 3600 & 1.10  \\
UM 461a              & Apr 12  & 3600 & 1.12  & IC 4662a             & Apr 13 & 2100 & 1.23  \\
J 1044+0353a         & Apr 13  & 3600 & 1.19  & J 1430+0027          & Apr 12 & 2400 & 1.67  \\
Tol 1214-277         & Apr 14  & 2400 & 1.09  & IC 4662c             & Apr 13 & 2100 & 1.23  \\
UM 570               & Apr 11  & 3600 & 1.13  & Pox 120b             & Apr 13 & 2700 & 1.05  \\
CGCG 032-017          & Apr 14  & 2400 & 1.24  & g2252292-171050      & Oct  9 & 1200 & 1.10  \\
Pox 108              & Apr 12  & 2700 & 1.47  & UM 461b              & Apr 12 & 3600 & 1.12   \\
UM 559               & Apr 14  & 2400 & 1.24  & Mrk 1236c            & Apr 14 & 1920 & 1.16  \\
SBS 0335-052 \#7     & Oct 7   & 2400 & 1.12  & J 1448-0110          & Apr 11 & 3600 & 1.22  \\
UM 442               & Apr 12  & 2400 & 1.18  & IC 4662b             & Apr 13 & 2100 & 1.23  \\
Mrk 1329             & Apr 12  & 3600 & 1.51  & J 2230-0006          & Oct  8 & 3600 & 1.30 \\
Mrk 1236a            & Apr 14  & 1920 & 1.16  & Pox 105              & Apr 11 & 3600 & 1.20 \\
Tol 2146-391         & Apr 13  & 2292 & 1.38  & UM 462d              & Apr 14 & 2400 & 1.12 \\
CGCG 007-025a         & Apr 12  & 3600 & 1.14  & UM 462c              & Apr 14 & 2400 & 1.12 \\
HS 2134+0400         & Oct 8   & 3600 & 1.27  & J 0248-0817b         & Oct  7 & 3600 & 1.07 \\
J 0014-0043          & Oct 9   & 2700 & 1.37  & CGCG 007-025b         & Apr 12 & 3600 & 1.14 \\
J 2324-0006          & Oct 7   & 3600 & 1.15  & J 1624-0022          & Apr 11 & 3600 & 1.18 \\
SBS 0335-052E \#3b   & Oct 9   & 2700 & 1.16  & UM 462e              & Apr 14 & 2400 & 1.12 \\
J 1253-0312          & Apr 13  & 2700 & 1.18  & J 0134-0038          & Oct  9 & 2400 & 1.18 \\
J 1201+0211          & Apr 14  & 2400 & 1.21  & SBS 0335-052W        & Oct  9 & 3600 & 1.10 \\
SBS 0335-052E \#3a   & Oct 9   & 2700 & 1.16  &  Tol65b              & Apr 13 & 3600 & 1.02 \\
Tol 65a              & Apr 13  & 3600 & 1.02  & SBS 0335-052 \#4,5   & Oct  9 & 2700 & 1.16 \\
UM 311               & Oct 8   & 1800 & 1.25  & UM 462b              & Apr 14 & 2400 & 1.12  \\
J 1304-0333          & Apr 13  & 1800 & 1.86  & CGCG 007-025d         & Apr 12 & 3600 & 1.14  \\
J 0248-0817a         & Oct 7   & 3600 & 1.07  & ESO 338-IG04         & Apr 12 & 1228 & 1.05  \\
Pox 4a               & Apr 12  & 3600 & 1.10  & Mrk 1236b            & Apr 14 & 1920 & 1.16  \\
Pox 120a             & Apr 13  & 2700 & 1.05  & J1414-0208           & Apr 13 & 3600 & 1.16  \\
HS 2236+1344b        & Oct 7   & 3600 & 1.38  & IC 4662d             & Apr 13 & 2100 & 1.23  \\
CGCG 007-025c         & Apr 12  & 3600 & 1.14  & J 0107+0103          & Oct  8 & 3200 & 1.17   \\
UM 462a              & Apr 14  & 2400 & 1.12  & g0405204-364859      & Oct  8 & 2400 & 1.03   \\
J 2302+0049          & Oct 8   & 2400 & 1.16  & J 1044+0353b         & Apr 13 & 3600 & 1.19   \\
Tol2138-405          & Apr 14  & 2260 & 1.35  & IC 4662e             & Apr 13 & 2100 & 1.23  \\
J 2104-0035N         & Oct 8   & 3600 & 1.16  &                      & &  \\
\hline
\end{tabular}

$^a${Names of the SDSS objects are given abridged  without seconds.} \\
$^b${Total exposure time.} \\
$^c${Air mass at the beginning of the observation.}
\end{table*}


\clearpage

\setcounter{figure}{2}
\begin{figure*}
\hspace*{1.0cm}\psfig{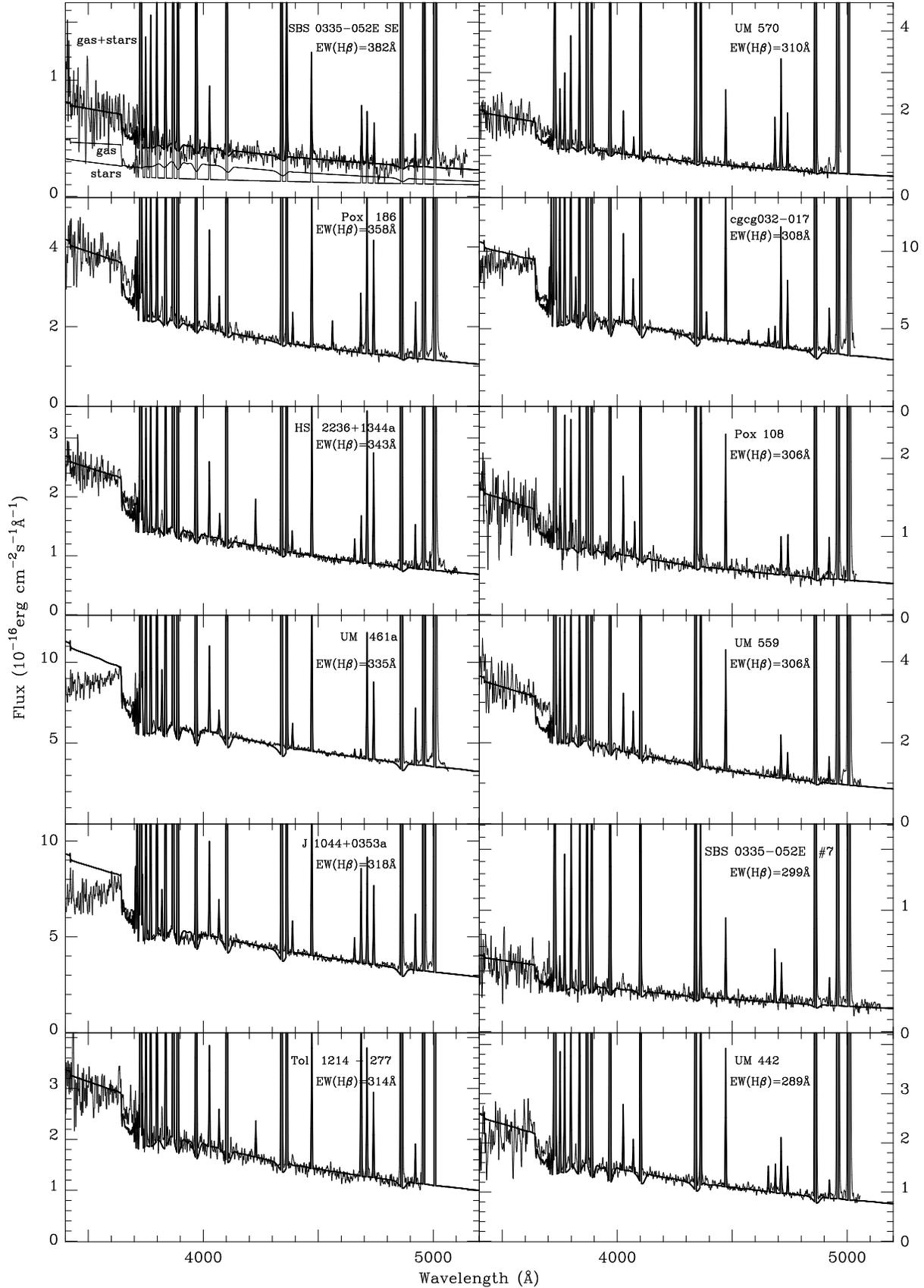}
\caption{Best fit model SEDs (thick solid lines) superposed by
the redshift- and extinction-corrected 3.6m   
spectra of 69 H {\sc ii} regions in 45 blue compact dwarf galaxies.
The separate contributions 
from the stellar and ionised gas components are shown only for SBS 0335--052E SE
by thin solid lines (see the labels in the upper left panel of the first page). 
Names of the SDSS objects are given in short form without seconds.  
\label{sp_1}}
\end{figure*}

\clearpage

\setcounter{figure}{2}
\begin{figure*}
\hspace*{1.0cm}\psfig{figure=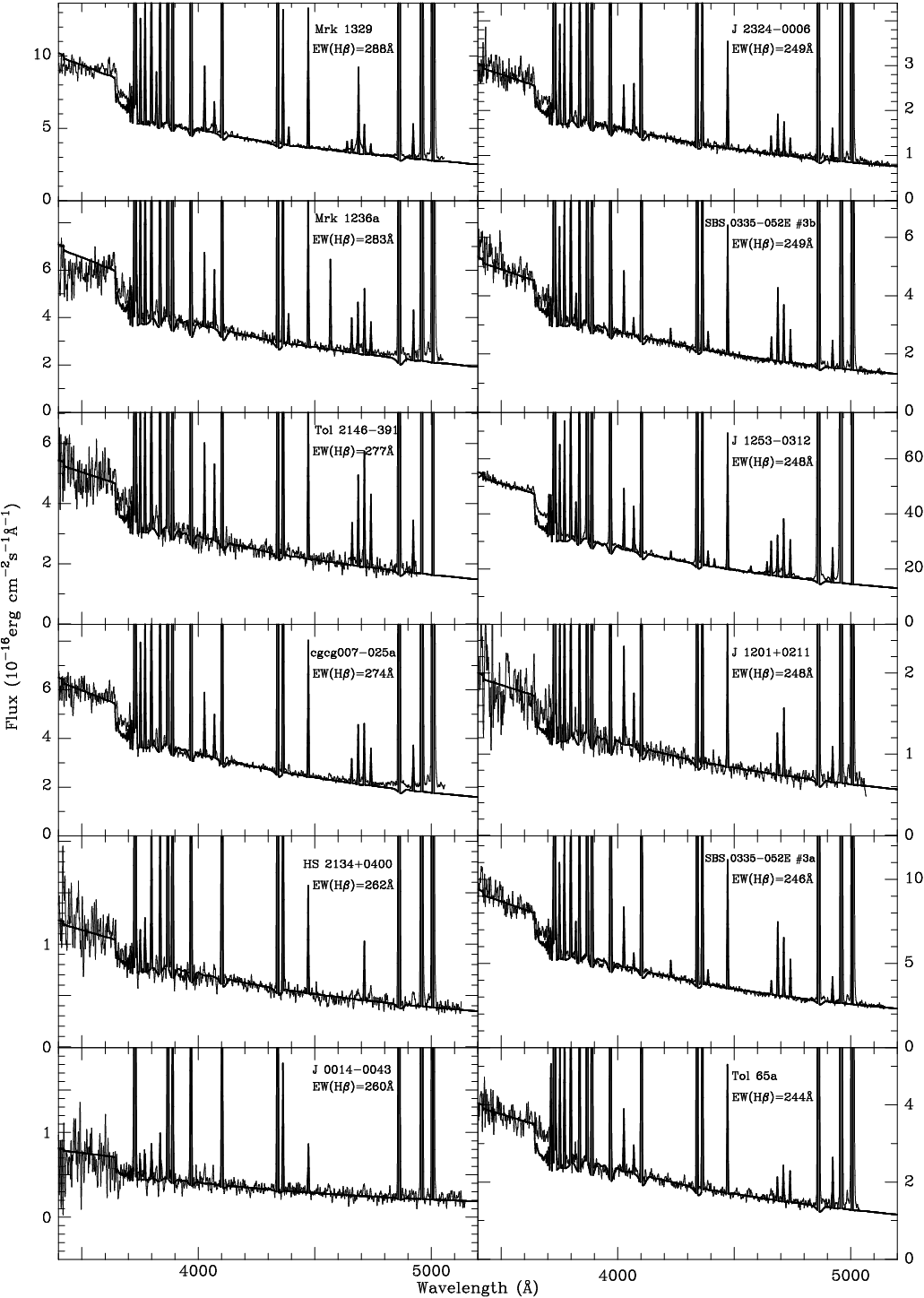,angle=0,width=16.cm,clip=}
\caption{--$Continued$}
    \label{sp_2}
\end{figure*}

\clearpage

\setcounter{figure}{2}
\begin{figure*}
\hspace*{1.0cm}\psfig{figure=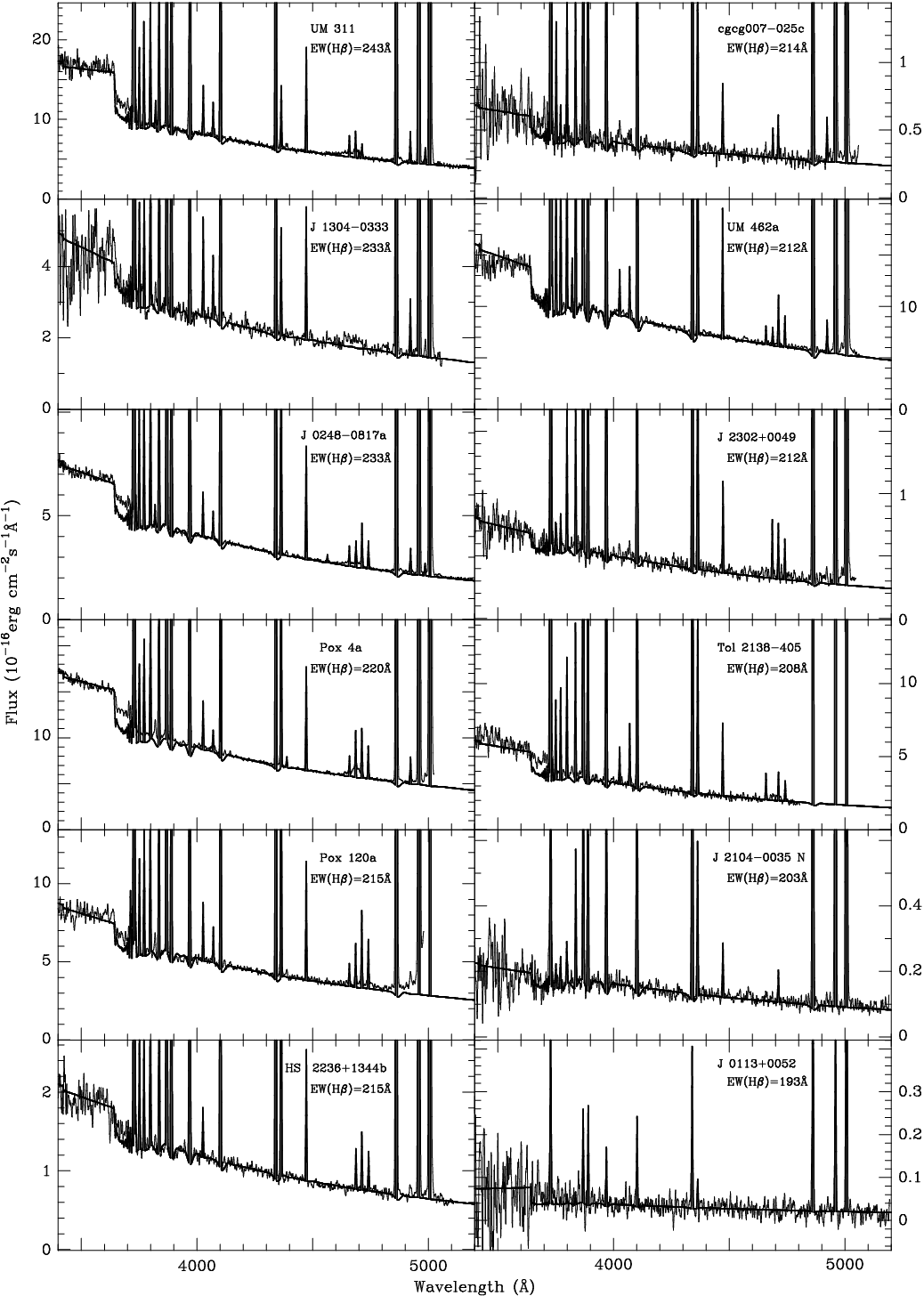,angle=0,width=16.cm,clip=}
\caption{--$Continued$}
    \label{sp_3}
\end{figure*}

\clearpage

\setcounter{figure}{2}
\begin{figure*}
\hspace*{1.0cm}\psfig{figure=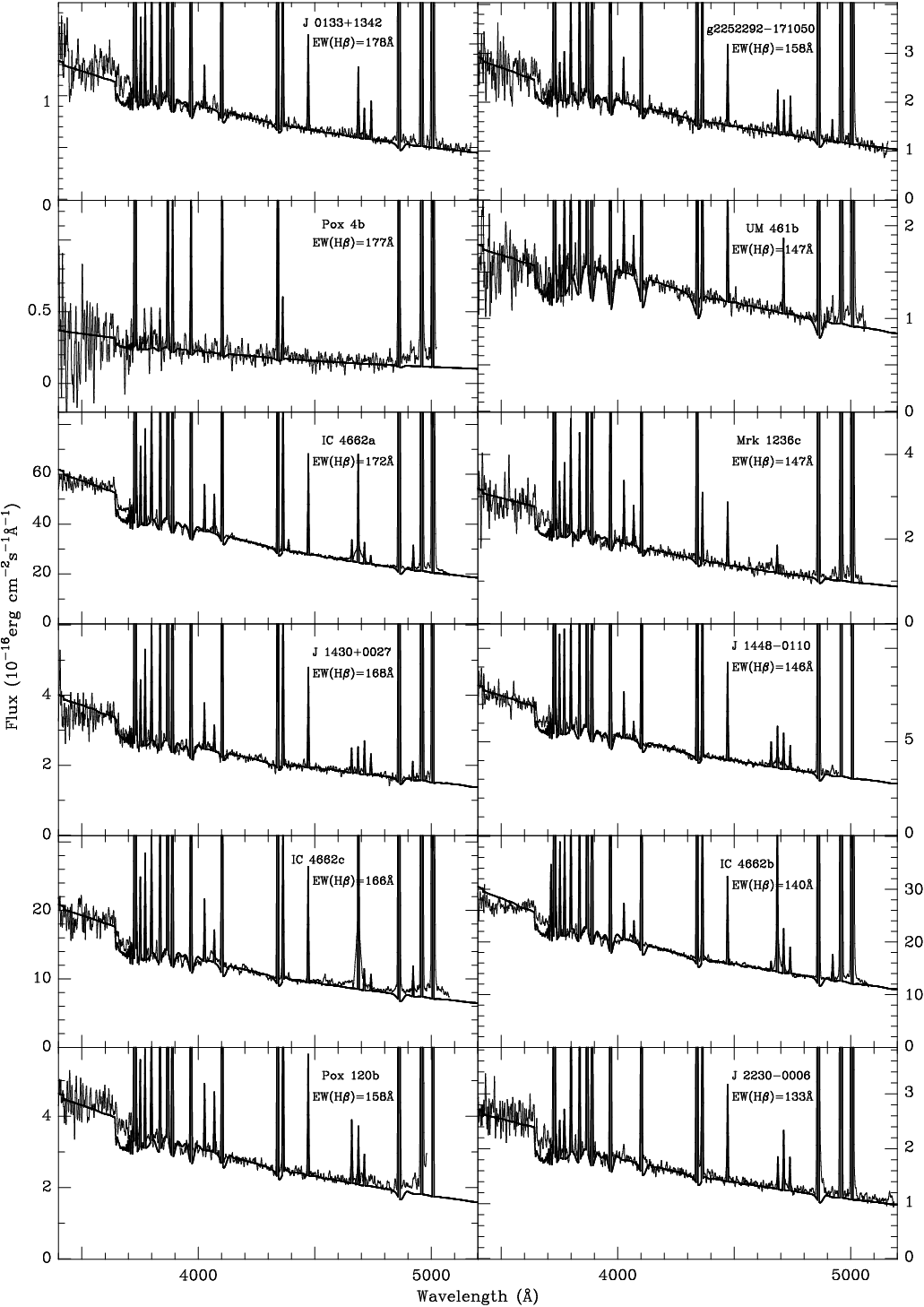,angle=0,width=16.cm,clip=}
\caption{--$Continued$}
    \label{sp_4}
\end{figure*}

\clearpage

\setcounter{figure}{2}
\begin{figure*}
\hspace*{1.0cm}\psfig{figure=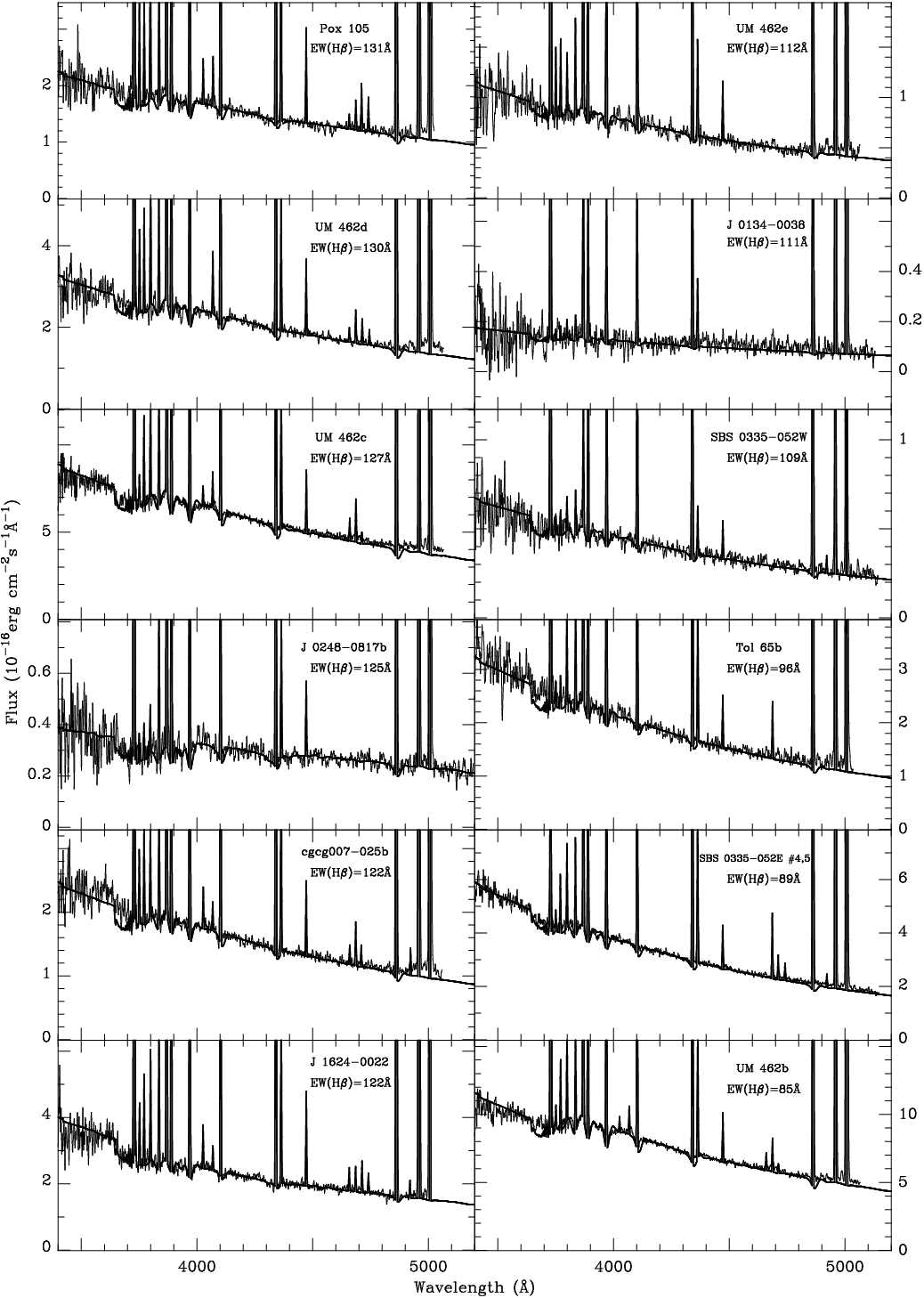,angle=0,width=16.cm,clip=}
\caption{--$Continued$}
    \label{sp_5}
\end{figure*}

\clearpage

\setcounter{figure}{2}
\begin{figure*}
\hspace*{1.0cm}\psfig{figure=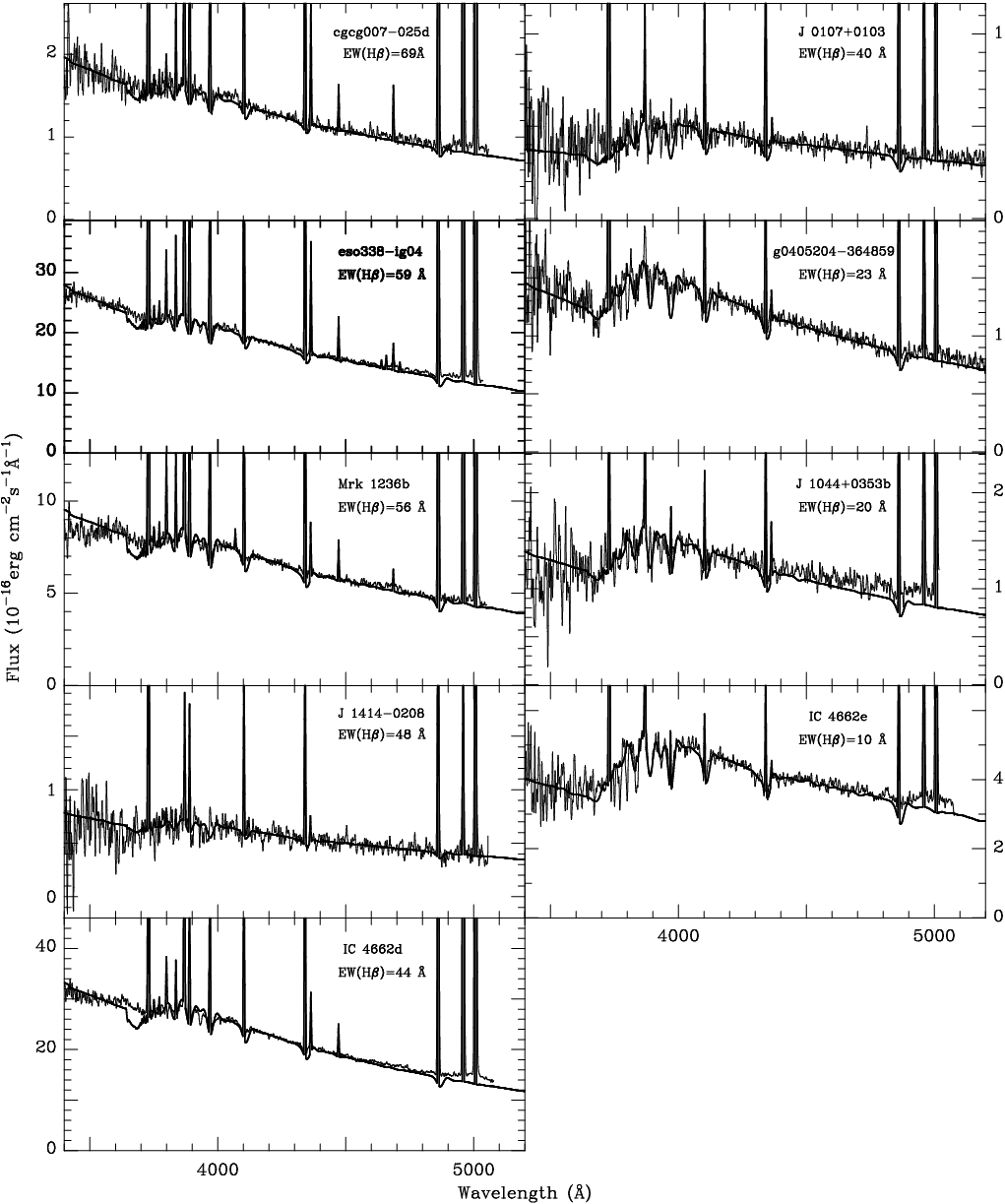,angle=0,width=16.cm,clip=}
\caption{--$Continued$}
    \label{sp_6}
\end{figure*}

\clearpage

\setcounter{table}{2}
  \begin{landscape}
  \begin{table}
  \centering
  \caption{Emission Line Intensities and  Equivalent Widths\label{t2_1}}
\begin{tabular}{lrrrcrrrcrrr} \\ \hline
   & \multicolumn{11}{c}{\sc Galaxy}\\              \cline{2-12} 
  &
 \multicolumn{3}{c}{Pox 186             }&&
 \multicolumn{3}{c}{HS 2236+1344a       }&&
 \multicolumn{3}{c}{UM 461a             } \\ 
  \cline{2-4} \cline{6-8} \cline{10-12}
  Ion
  &$F$($\lambda$)/$F$(H$\beta$)
  &$I$($\lambda$)/$I$(H$\beta$)
  &EW$^a$&
  &$F$($\lambda$)/$F$(H$\beta$)
  &$I$($\lambda$)/$I$(H$\beta$)
  &EW$^a$&
  &$F$($\lambda$)/$F$(H$\beta$)
  &$I$($\lambda$)/$I$(H$\beta$)
  &EW$^a$ \\ \hline
3704 H16                          &   0.95 $\pm$   0.13 &   1.09 $\pm$   0.65 &   1.7 & &   ...~~~~~~ &   ...~~~~~~ &  ...~~~& &   ...~~~~~~ &   ...~~~~~~ &  ...~~~\\
3712 H15                          &   0.97 $\pm$   0.14 &   1.11 $\pm$   0.61 &   1.9 & &   ...~~~~~~ &   ...~~~~~~ &  ...~~~& &   ...~~~~~~ &   ...~~~~~~ &  ...~~~\\
3727 [O {\sc ii}]                 &  33.03 $\pm$   0.53 &  34.94 $\pm$   0.58 &  68.9 & &  25.63 $\pm$   0.42 &  28.28 $\pm$   0.49 &  49.3 & &  35.79 $\pm$   0.52 &  41.33 $\pm$   0.64 &  93.5 \\
3750 H12                          &   3.01 $\pm$   0.16 &   3.25 $\pm$   0.22 &   6.8 & &   2.45 $\pm$   0.16 &   2.81 $\pm$   0.25 &   5.3 & &   2.23 $\pm$   0.06 &   4.25 $\pm$   0.14 &   6.2 \\
\hline
  \end{tabular}
  \end{table}
\clearpage

\setcounter{table}{2}

  \begin{table}
  \centering
  \caption{\sl ---Continued.}
  \begin{tabular}{lrrrcrrrcrrr} \\ \hline
   & \multicolumn{11}{c}{\sc Galaxy}\\              \cline{2-12} 
  &
 \multicolumn{3}{c}{Tol 1214-277        }&&
 \multicolumn{3}{c}{UM 570              }&&
 \multicolumn{3}{c}{CGCG 032-017        } \\ 
  \cline{2-4} \cline{6-8} \cline{10-12}
  Ion
  &$F$($\lambda$)/$F$(H$\beta$)
  &$I$($\lambda$)/$I$(H$\beta$)
  &EW$^a$&
  &$F$($\lambda$)/$F$(H$\beta$)
  &$I$($\lambda$)/$I$(H$\beta$)
  &EW$^a$&
  &$F$($\lambda$)/$F$(H$\beta$)
  &$I$($\lambda$)/$I$(H$\beta$)
  &EW$^a$ \\ \hline
3704 H16                          &   ...~~~~~~ &   ...~~~~~~ &  ...~~~& &   ...~~~~~~ &   ...~~~~~~ &  ...~~~& &   0.78 $\pm$   0.11 &   2.49 $\pm$   0.68 &   1.7 \\
3712 H15                          &   ...~~~~~~ &   ...~~~~~~ &  ...~~~& &   ...~~~~~~ &   ...~~~~~~ &  ...~~~& &   0.49 $\pm$   0.16 &   2.13 $\pm$   1.47 &   1.1 \\
3727 [O {\sc ii}]                 &  23.73 $\pm$   0.43 &  26.01 $\pm$   0.49 &  42.5 & &  19.48 $\pm$   0.48 &  21.43 $\pm$   0.54 &  26.0 & &  66.80 $\pm$   0.97 &  80.17 $\pm$   1.24 & 155.6 \\
3750 H12                          &   2.76 $\pm$   0.19 &   4.22 $\pm$   0.35 &   5.9 & &   2.74 $\pm$   0.28 &   3.01 $\pm$   0.45 &   4.6 & &   2.05 $\pm$   0.10 &   3.73 $\pm$   0.22 &   5.3 \\
\hline
  \end{tabular}
  \end{table}
\clearpage

\setcounter{table}{2}

  \begin{table}
  \centering
  \caption{\sl ---Continued.}
  \begin{tabular}{lrrrcrrrcrrr} \\ \hline
   & \multicolumn{11}{c}{\sc Galaxy}\\              \cline{2-12} 
  &
 \multicolumn{3}{c}{Pox 108             }&&
 \multicolumn{3}{c}{UM 559              }&&
 \multicolumn{3}{c}{UM 442              } \\ 
  \cline{2-4} \cline{6-8} \cline{10-12}
  Ion
  &$F$($\lambda$)/$F$(H$\beta$)
  &$I$($\lambda$)/$I$(H$\beta$)
  &EW$^a$&
  &$F$($\lambda$)/$F$(H$\beta$)
  &$I$($\lambda$)/$I$(H$\beta$)
  &EW$^a$&
  &$F$($\lambda$)/$F$(H$\beta$)
  &$I$($\lambda$)/$I$(H$\beta$)
  &EW$^a$ \\ \hline
3727 [O {\sc ii}]                 &  80.28 $\pm$   1.35 &  84.51 $\pm$   1.49 & 112.2 & &  82.22 $\pm$   1.25 &  86.10 $\pm$   1.37 & 125.5 & &  64.90 $\pm$   1.04 &  71.35 $\pm$   1.20 & 119.7 \\
3750 H12                          &   ...~~~~~~ &   ...~~~~~~ &  ...~~~& &   3.74 $\pm$   0.26 &   3.92 $\pm$   0.34 &   6.4 & &   2.86 $\pm$   0.30 &   3.85 $\pm$   0.50 &   5.8 \\
3771 H11                          &   4.23 $\pm$   0.36 &   4.75 $\pm$   0.53 &   8.2 & &   4.45 $\pm$   0.23 &   4.65 $\pm$   0.32 &   7.3 & &   3.37 $\pm$   0.30 &   4.46 $\pm$   0.51 &   6.2 \\
3798 H10                          &   4.01 $\pm$   0.32 &   4.54 $\pm$   0.53 &   6.9 & &   5.59 $\pm$   0.25 &   5.83 $\pm$   0.33 &   9.5 & &   3.69 $\pm$   0.21 &   4.77 $\pm$   0.41 &   7.0 \\
\hline
  \end{tabular}
  \end{table}
\clearpage

\setcounter{table}{2}

  \begin{table}
  \centering
  \caption{\sl ---Continued.}
  \begin{tabular}{lrrrcrrrcrrr} \\ \hline
   & \multicolumn{11}{c}{\sc Galaxy}\\              \cline{2-12} 
  &
 \multicolumn{3}{c}{Mrk 1329            }&&
 \multicolumn{3}{c}{Mrk 1236a           }&&
 \multicolumn{3}{c}{Tol 2146-391        } \\ 
  \cline{2-4} \cline{6-8} \cline{10-12}
  Ion
  &$F$($\lambda$)/$F$(H$\beta$)
  &$I$($\lambda$)/$I$(H$\beta$)
  &EW$^a$&
  &$F$($\lambda$)/$F$(H$\beta$)
  &$I$($\lambda$)/$I$(H$\beta$)
  &EW$^a$&
  &$F$($\lambda$)/$F$(H$\beta$)
  &$I$($\lambda$)/$I$(H$\beta$)
  &EW$^a$ \\ \hline
3727 [O {\sc ii}]                 &  96.34 $\pm$   1.38 &  98.03 $\pm$   1.48 & 157.1 & & 119.88 $\pm$   1.75 & 131.00 $\pm$   2.04 & 213.5 & &  66.08 $\pm$   1.00 &  75.07 $\pm$   1.20 &  99.1 \\
3750 H12                          &   2.52 $\pm$   0.08 &   3.22 $\pm$   0.13 &   4.3 & &   2.65 $\pm$   0.15 &   4.62 $\pm$   0.33 &   5.0 & &   2.93 $\pm$   0.21 &   3.32 $\pm$   0.29 &   5.6 \\
3771 H11                          &   3.41 $\pm$   0.09 &   4.11 $\pm$   0.14 &   5.9 & &   2.78 $\pm$   0.15 &   4.71 $\pm$   0.31 &   5.4 & &   3.28 $\pm$   0.17 &   3.71 $\pm$   0.26 &   6.1 \\
3798 H10                          &   4.55 $\pm$   0.10 &   5.25 $\pm$   0.14 &   8.1 & &   4.23 $\pm$   0.15 &   6.20 $\pm$   0.28 &   8.6 & &   4.83 $\pm$   0.19 &   5.44 $\pm$   0.28 &   8.7 \\
\hline
  \end{tabular}
  \end{table}
\clearpage

\setcounter{table}{2}

  \begin{table}
  \centering
  \caption{\sl ---Continued.}
  \begin{tabular}{lrrrcrrrcrrr} \\ \hline
   & \multicolumn{11}{c}{\sc Galaxy}\\              \cline{2-12} 
  &
 \multicolumn{3}{c}{CGCG 007-025a       }&&
 \multicolumn{3}{c}{HS 2134+0400        }&&
 \multicolumn{3}{c}{J 00143498-00435203 } \\ 
  \cline{2-4} \cline{6-8} \cline{10-12}
  Ion
  &$F$($\lambda$)/$F$(H$\beta$)
  &$I$($\lambda$)/$I$(H$\beta$)
  &EW$^a$&
  &$F$($\lambda$)/$F$(H$\beta$)
  &$I$($\lambda$)/$I$(H$\beta$)
  &EW$^a$&
  &$F$($\lambda$)/$F$(H$\beta$)
  &$I$($\lambda$)/$I$(H$\beta$)
  &EW$^a$ \\ \hline
3727 [O {\sc ii}]                 &  75.86 $\pm$   1.11 &  87.71 $\pm$   1.35 & 128.6 & &  38.53 $\pm$   0.87 &  46.00 $\pm$   1.07 &  57.6 & &  97.03 $\pm$   2.20 & 103.64 $\pm$   2.48 & 133.5 \\
3750 H12                          &   2.52 $\pm$   0.12 &   2.94 $\pm$   0.20 &   4.4 & &   ...~~~~~~ &   ...~~~~~~ &  ...~~~& &   ...~~~~~~ &   ...~~~~~~ &  ...~~~\\
3771 H11                          &   3.38 $\pm$   0.13 &   3.92 $\pm$   0.21 &   6.3 & &   ...~~~~~~ &   ...~~~~~~ &  ...~~~& &   ...~~~~~~ &   ...~~~~~~ &  ...~~~\\
3798 H10                          &   4.67 $\pm$   0.13 &   5.37 $\pm$   0.20 &   8.6 & &   5.50 $\pm$   0.52 &   6.49 $\pm$   0.78 &   9.6 & &   ...~~~~~~ &   ...~~~~~~ &  ...~~~\\
\hline
  \end{tabular}
  \end{table}
\clearpage

\setcounter{table}{2}

  \begin{table}
  \centering
  \caption{\sl ---Continued.}
  \begin{tabular}{lrrrcrrrcrrr} \\ \hline
   & \multicolumn{11}{c}{\sc Galaxy}\\              \cline{2-12} 
  &
 \multicolumn{3}{c}{J 23242034-00062507 }&&
 \multicolumn{3}{c}{J 12530596-03125894 }&&
 \multicolumn{3}{c}{Tol 65a             } \\ 
  \cline{2-4} \cline{6-8} \cline{10-12}
  Ion
  &$F$($\lambda$)/$F$(H$\beta$)
  &$I$($\lambda$)/$I$(H$\beta$)
  &EW$^a$&
  &$F$($\lambda$)/$F$(H$\beta$)
  &$I$($\lambda$)/$I$(H$\beta$)
  &EW$^a$&
  &$F$($\lambda$)/$F$(H$\beta$)
  &$I$($\lambda$)/$I$(H$\beta$)
  &EW$^a$ \\ \hline
3704 H16                          &   ...~~~~~~ &   ...~~~~~~ &  ...~~~& &   0.72 $\pm$   0.05 &   0.93 $\pm$   0.50 &   1.0 & &   ...~~~~~~ &   ...~~~~~~ &  ...~~~\\
3712 H15                          &   ...~~~~~~ &   ...~~~~~~ &  ...~~~& &   0.64 $\pm$   0.05 &   0.82 $\pm$   0.56 &   0.9 & &   1.51 $\pm$   0.19 &   2.42 $\pm$   0.81 &   2.2 \\
3727 [O {\sc ii}]                 & 124.53 $\pm$   1.83 & 143.89 $\pm$   2.23 & 162.6 & &  69.73 $\pm$   1.00 &  89.68 $\pm$   1.35 & 101.3 & &  64.46 $\pm$   0.98 &  68.88 $\pm$   1.10 &  93.8 \\
3750 H12                          &   3.06 $\pm$   0.20 &   3.67 $\pm$   0.33 &   4.8 & &   2.28 $\pm$   0.06 &   2.92 $\pm$   0.10 &   3.5 & &   2.55 $\pm$   0.17 &   3.50 $\pm$   0.31 &   3.8 \\
\hline
  \end{tabular}
  \end{table}
\clearpage

\setcounter{table}{2}

  \begin{table}
  \centering
  \caption{\sl ---Continued.}
  \begin{tabular}{lrrrcrrrcrrr} \\ \hline
   & \multicolumn{11}{c}{\sc Galaxy}\\              \cline{2-12} 
  &
 \multicolumn{3}{c}{UM311               }&&
 \multicolumn{3}{c}{J 13043227-03332212 }&&
 \multicolumn{3}{c}{J 02481595-08171651a} \\ 
  \cline{2-4} \cline{6-8} \cline{10-12}
  Ion
  &$F$($\lambda$)/$F$(H$\beta$)
  &$I$($\lambda$)/$I$(H$\beta$)
  &EW$^a$&
  &$F$($\lambda$)/$F$(H$\beta$)
  &$I$($\lambda$)/$I$(H$\beta$)
  &EW$^a$&
  &$F$($\lambda$)/$F$(H$\beta$)
  &$I$($\lambda$)/$I$(H$\beta$)
  &EW$^a$ \\ \hline
3727 [O {\sc ii}]                 & 149.64 $\pm$   2.15 & 174.35 $\pm$   2.65 & 196.1 & & 173.38 $\pm$   2.64 & 187.74 $\pm$   3.01 & 209.9 & &  84.40 $\pm$   1.22 &  86.43 $\pm$   1.32 & 101.1 \\
3750 H12                          &   2.21 $\pm$   0.15 &   3.16 $\pm$   0.27 &   3.2 & &   2.63 $\pm$   0.28 &   3.21 $\pm$   0.60 &   3.4 & &   2.67 $\pm$   0.11 &   4.25 $\pm$   0.21 &   3.4 \\
3771 H11                          &   3.14 $\pm$   0.16 &   4.22 $\pm$   0.26 &   4.6 & &   3.05 $\pm$   0.28 &   3.64 $\pm$   0.57 &   4.2 & &   3.37 $\pm$   0.11 &   4.93 $\pm$   0.20 &   4.4 \\
3798 H10                          &   3.90 $\pm$   0.14 &   5.10 $\pm$   0.24 &   5.6 & &   5.26 $\pm$   0.34 &   5.99 $\pm$   0.58 &   7.8 & &   4.45 $\pm$   0.11 &   6.04 $\pm$   0.18 &   5.8 \\
\hline
  \end{tabular}
  \end{table}
\clearpage

\setcounter{table}{2}

  \begin{table}
  \centering
  \caption{\sl ---Continued.}
  \begin{tabular}{lrrrcrrrcrrr} \\ \hline
   & \multicolumn{11}{c}{\sc Galaxy}\\              \cline{2-12} 
  &
 \multicolumn{3}{c}{Pox 4a              }&&
 \multicolumn{3}{c}{Pox 120a            }&&
 \multicolumn{3}{c}{HS 2236+1344b       } \\ 
  \cline{2-4} \cline{6-8} \cline{10-12}
  Ion
  &$F$($\lambda$)/$F$(H$\beta$)
  &$I$($\lambda$)/$I$(H$\beta$)
  &EW$^a$&
  &$F$($\lambda$)/$F$(H$\beta$)
  &$I$($\lambda$)/$I$(H$\beta$)
  &EW$^a$&
  &$F$($\lambda$)/$F$(H$\beta$)
  &$I$($\lambda$)/$I$(H$\beta$)
  &EW$^a$ \\ \hline
3712 H15                          &   ...~~~~~~ &   ...~~~~~~ &  ...~~~& &   1.00 $\pm$   0.14 &   1.63 $\pm$   0.90 &   1.4 & &   ...~~~~~~ &   ...~~~~~~ &  ...~~~\\
3727 [O {\sc ii}]                 &  71.17 $\pm$   1.02 &  75.29 $\pm$   1.14 &  80.8 & &  56.83 $\pm$   0.86 &  65.79 $\pm$   1.04 &  80.9 & &  39.91 $\pm$   0.72 &  46.48 $\pm$   0.87 &  49.1 \\
3750 H12                          &   2.56 $\pm$   0.07 &   2.80 $\pm$   0.11 &   3.1 & &   2.39 $\pm$   0.13 &   3.16 $\pm$   0.22 &   3.7 & &   3.43 $\pm$   0.29 &   3.99 $\pm$   0.46 &   4.7 \\
3771 H11                          &   3.45 $\pm$   0.08 &   3.72 $\pm$   0.11 &   4.2 & &   3.35 $\pm$   0.14 &   4.27 $\pm$   0.24 &   5.1 & &   3.26 $\pm$   0.25 &   3.78 $\pm$   0.43 &   4.3 \\
\hline
  \end{tabular}
  \end{table}
\clearpage

\setcounter{table}{2}

  \begin{table}
  \centering
  \caption{\sl ---Continued.}
  \begin{tabular}{lrrrcrrrcrrr} \\ \hline
   & \multicolumn{11}{c}{\sc Galaxy}\\              \cline{2-12} 
  &
 \multicolumn{3}{c}{CGCG 007-025c       }&&
 \multicolumn{3}{c}{UM 462a             }&&
 \multicolumn{3}{c}{Tol 2138-405        } \\ 
  \cline{2-4} \cline{6-8} \cline{10-12}
  Ion
  &$F$($\lambda$)/$F$(H$\beta$)
  &$I$($\lambda$)/$I$(H$\beta$)
  &EW$^a$&
  &$F$($\lambda$)/$F$(H$\beta$)
  &$I$($\lambda$)/$I$(H$\beta$)
  &EW$^a$&
  &$F$($\lambda$)/$F$(H$\beta$)
  &$I$($\lambda$)/$I$(H$\beta$)
  &EW$^a$ \\ \hline
3727 [O {\sc ii}]                 & 112.66 $\pm$   1.91 & 114.27 $\pm$   2.03 & 176.4 & &  88.85 $\pm$   1.31 & 127.27 $\pm$   1.99 & 152.2 & & 108.82 $\pm$   1.67 & 143.48 $\pm$   2.32 & 145.9 \\
3750 H12                          &   6.26 $\pm$   0.67 &   6.35 $\pm$   0.87 &  10.9 & &   2.06 $\pm$   0.16 &   4.43 $\pm$   0.40 &   3.8 & &   3.10 $\pm$   0.26 &   4.05 $\pm$   0.42 &   4.9 \\
3771 H11                          &   ...~~~~~~ &   ...~~~~~~ &  ...~~~& &   1.98 $\pm$   0.12 &   4.41 $\pm$   0.34 &   3.4 & &   3.62 $\pm$   0.22 &   4.72 $\pm$   0.38 &   5.6 \\
3798 H10                          &   7.37 $\pm$   0.56 &   7.47 $\pm$   0.77 &  13.4 & &   2.57 $\pm$   0.12 &   5.20 $\pm$   0.31 &   4.4 & &   4.77 $\pm$   0.23 &   6.17 $\pm$   0.39 &   7.4 \\
\hline
  \end{tabular}
  \end{table}
\clearpage

\setcounter{table}{2}

  \begin{table}
  \centering
  \caption{\sl ---Continued.}
  \begin{tabular}{lrrrcrrrcrrr} \\ \hline
   & \multicolumn{11}{c}{\sc Galaxy}\\              \cline{2-12} 
  &
 \multicolumn{3}{c}{Pox 4b              }&&
 \multicolumn{3}{c}{IC 4662a            }&&
 \multicolumn{3}{c}{J 14305350+00274635 } \\ 
  \cline{2-4} \cline{6-8} \cline{10-12}
  Ion
  &$F$($\lambda$)/$F$(H$\beta$)
  &$I$($\lambda$)/$I$(H$\beta$)
  &EW$^a$&
  &$F$($\lambda$)/$F$(H$\beta$)
  &$I$($\lambda$)/$I$(H$\beta$)
  &EW$^a$&
  &$F$($\lambda$)/$F$(H$\beta$)
  &$I$($\lambda$)/$I$(H$\beta$)
  &EW$^a$ \\ \hline
3727 [O {\sc ii}]                 & 194.71 $\pm$   4.27 & 192.90 $\pm$   4.51 & 212.0 & & 106.93 $\pm$   1.53 & 146.20 $\pm$   2.22 & 124.3 & & 126.12 $\pm$   1.91 & 140.27 $\pm$   2.24 & 142.7 \\
3750 H12                          &   ...~~~~~~ &   ...~~~~~~ &  ...~~~& &   1.77 $\pm$   0.10 &   2.90 $\pm$   0.19 &   2.2 & &   2.48 $\pm$   0.24 &   2.75 $\pm$   0.36 &   3.0 \\
3771 H11                          &   ...~~~~~~ &   ...~~~~~~ &  ...~~~& &   2.29 $\pm$   0.09 &   3.59 $\pm$   0.17 &   2.8 & &   3.66 $\pm$   0.24 &   4.05 $\pm$   0.37 &   4.4 \\
3798 H10                          &   ...~~~~~~ &   ...~~~~~~ &  ...~~~& &   3.34 $\pm$   0.10 &   4.96 $\pm$   0.18 &   4.1 & &   4.51 $\pm$   0.26 &   4.98 $\pm$   0.38 &   5.5 \\
\hline
  \end{tabular}
  \end{table}
\clearpage

\setcounter{table}{2}

  \begin{table}
  \centering
  \caption{\sl ---Continued.}
  \begin{tabular}{lrrrcrrrcrrr} \\ \hline
   & \multicolumn{11}{c}{\sc Galaxy}\\              \cline{2-12} 
  &
 \multicolumn{3}{c}{IC 4662c            }&&
 \multicolumn{3}{c}{Pox 120b            }&&
 \multicolumn{3}{c}{g2252292-171050     } \\ 
  \cline{2-4} \cline{6-8} \cline{10-12}
  Ion
  &$F$($\lambda$)/$F$(H$\beta$)
  &$I$($\lambda$)/$I$(H$\beta$)
  &EW$^a$&
  &$F$($\lambda$)/$F$(H$\beta$)
  &$I$($\lambda$)/$I$(H$\beta$)
  &EW$^a$&
  &$F$($\lambda$)/$F$(H$\beta$)
  &$I$($\lambda$)/$I$(H$\beta$)
  &EW$^a$ \\ \hline
3727 [O {\sc ii}]                 & 130.24 $\pm$   1.89 & 164.56 $\pm$   2.52 & 166.6 & & 112.03 $\pm$   1.70 & 124.15 $\pm$   1.99 & 108.1 & &  65.43 $\pm$   1.20 &  69.92 $\pm$   1.34 &  61.7 \\
3750 H12                          &   2.49 $\pm$   0.16 &   3.13 $\pm$   0.28 &   3.6 & &   2.72 $\pm$   0.25 &   3.01 $\pm$   0.37 &   2.8 & &   ...~~~~~~ &   ...~~~~~~ &  ...~~~\\
3771 H11                          &   3.24 $\pm$   0.19 &   4.06 $\pm$   0.31 &   4.8 & &   3.09 $\pm$   0.25 &   3.41 $\pm$   0.36 &   3.3 & &   ...~~~~~~ &   ...~~~~~~ &  ...~~~\\
3798 H10                          &   3.85 $\pm$   0.15 &   4.79 $\pm$   0.27 &   5.4 & &   4.45 $\pm$   0.25 &   4.89 $\pm$   0.36 &   4.7 & &   3.98 $\pm$   0.53 &   4.39 $\pm$   0.80 &   4.2 \\
\hline
  \end{tabular}
  \end{table}
\clearpage

\setcounter{table}{2}

  \begin{table}
  \centering
  \caption{\sl ---Continued.}
  \begin{tabular}{lrrrcrrrcrrr} \\ \hline
   & \multicolumn{11}{c}{\sc Galaxy}\\              \cline{2-12} 
  &
 \multicolumn{3}{c}{UM 461b             }&&
 \multicolumn{3}{c}{Mrk 1236c           }&&
 \multicolumn{3}{c}{J 14480536-01105771 } \\ 
  \cline{2-4} \cline{6-8} \cline{10-12}
  Ion
  &$F$($\lambda$)/$F$(H$\beta$)
  &$I$($\lambda$)/$I$(H$\beta$)
  &EW$^a$&
  &$F$($\lambda$)/$F$(H$\beta$)
  &$I$($\lambda$)/$I$(H$\beta$)
  &EW$^a$&
  &$F$($\lambda$)/$F$(H$\beta$)
  &$I$($\lambda$)/$I$(H$\beta$)
  &EW$^a$ \\ \hline
3727 [O {\sc ii}]                 &  63.86 $\pm$   1.10 &  71.04 $\pm$   1.30 &  69.6 & & 241.49 $\pm$   3.71 & 241.49 $\pm$   3.91 & 179.2 & &  84.27 $\pm$   1.25 &  97.26 $\pm$   1.53 &  78.4 \\
3750 H12                          &   ...~~~~~~ &   ...~~~~~~ &  ...~~~& &   3.14 $\pm$   0.53 &   3.14 $\pm$   0.79 &   2.5 & &   2.64 $\pm$   0.23 &   3.97 $\pm$   0.42 &   2.6 \\
3771 H11                          &   2.95 $\pm$   0.29 &   5.39 $\pm$   0.67 &   4.0 & &   4.27 $\pm$   0.57 &   4.27 $\pm$   0.79 &   3.6 & &   2.86 $\pm$   0.19 &   4.20 $\pm$   0.36 &   2.9 \\
3798 H10                          &   4.12 $\pm$   0.35 &   6.70 $\pm$   0.70 &   5.5 & &   6.53 $\pm$   0.56 &   6.53 $\pm$   0.77 &   5.6 & &   3.71 $\pm$   0.17 &   5.15 $\pm$   0.33 &   3.7 \\
\hline
  \end{tabular}
  \end{table}
\clearpage

\setcounter{table}{2}

  \begin{table}
  \centering
  \caption{\sl ---Continued.}
  \begin{tabular}{lrrrcrrrcrrr} \\ \hline
   & \multicolumn{11}{c}{\sc Galaxy}\\              \cline{2-12} 
  &
 \multicolumn{3}{c}{IC 4662b            }&&
 \multicolumn{3}{c}{Pox 105             }&&
 \multicolumn{3}{c}{UM 462d             } \\ 
  \cline{2-4} \cline{6-8} \cline{10-12}
  Ion
  &$F$($\lambda$)/$F$(H$\beta$)
  &$I$($\lambda$)/$I$(H$\beta$)
  &EW$^a$&
  &$F$($\lambda$)/$F$(H$\beta$)
  &$I$($\lambda$)/$I$(H$\beta$)
  &EW$^a$&
  &$F$($\lambda$)/$F$(H$\beta$)
  &$I$($\lambda$)/$I$(H$\beta$)
  &EW$^a$ \\ \hline
3712 H15                          &   0.97 $\pm$   0.12 &   2.76 $\pm$   1.06 &   0.9 & &   ...~~~~~~ &   ...~~~~~~ &  ...~~~& &   ...~~~~~~ &   ...~~~~~~ &  ...~~~\\
3727 [O {\sc ii}]                 &  70.08 $\pm$   1.02 &  76.86 $\pm$   1.19 &  66.1 & & 108.10 $\pm$   1.68 & 109.99 $\pm$   1.82 &  98.1 & & 188.51 $\pm$   2.92 & 219.94 $\pm$   3.63 & 173.6 \\
3750 H12                          &   2.05 $\pm$   0.11 &   3.90 $\pm$   0.26 &   2.0 & &   3.31 $\pm$   0.32 &   4.84 $\pm$   0.60 &   3.3 & &   2.29 $\pm$   0.41 &   4.13 $\pm$   0.91 &   2.1 \\
3771 H11                          &   2.89 $\pm$   0.13 &   4.77 $\pm$   0.27 &   2.8 & &   3.72 $\pm$   0.50 &   5.31 $\pm$   0.82 &   3.6 & &   3.32 $\pm$   0.42 &   5.20 $\pm$   0.78 &   3.3 \\
\hline
  \end{tabular}
  \end{table}
\clearpage

\setcounter{table}{2}

  \begin{table}
  \centering
  \caption{\sl ---Continued.}
  \begin{tabular}{lrrrcrrrcrrr} \\ \hline
   & \multicolumn{11}{c}{\sc Galaxy}\\              \cline{2-12} 
  &
 \multicolumn{3}{c}{UM 462c             }&&
 \multicolumn{3}{c}{J 02481595-08171651b}&&
 \multicolumn{3}{c}{cgcg007-025b        } \\ 
  \cline{2-4} \cline{6-8} \cline{10-12}
  Ion
  &$F$($\lambda$)/$F$(H$\beta$)
  &$I$($\lambda$)/$I$(H$\beta$)
  &EW$^a$&
  &$F$($\lambda$)/$F$(H$\beta$)
  &$I$($\lambda$)/$I$(H$\beta$)
  &EW$^a$&
  &$F$($\lambda$)/$F$(H$\beta$)
  &$I$($\lambda$)/$I$(H$\beta$)
  &EW$^a$ \\ \hline
3727 [O {\sc ii}]                 & 191.76 $\pm$   2.83 & 191.76 $\pm$   2.98 & 120.3 & & 124.64 $\pm$   2.39 & 120.54 $\pm$   2.50 & 131.1 & & 118.66 $\pm$   1.83 & 123.13 $\pm$   2.00 &  82.2 \\
3750 H12                          &   3.76 $\pm$   0.30 &   3.76 $\pm$   0.38 &   2.4 & &   ...~~~~~~ &   ...~~~~~~ &  ...~~~& &   3.18 $\pm$   0.32 &   3.64 $\pm$   0.55 &   2.4 \\
3771 H11                          &   3.54 $\pm$   0.26 &   3.54 $\pm$   0.35 &   2.2 & &   ...~~~~~~ &   ...~~~~~~ &  ...~~~& &   4.74 $\pm$   0.40 &   5.25 $\pm$   0.59 &   3.6 \\
3798 H10                          &   5.30 $\pm$   0.23 &   5.30 $\pm$   0.33 &   3.4 & &   ...~~~~~~ &   ...~~~~~~ &  ...~~~& &   4.06 $\pm$   0.37 &   4.55 $\pm$   0.59 &   3.0 \\
\hline
  \end{tabular}
  \end{table}
\clearpage

\setcounter{table}{2}

  \begin{table}
  \centering
  \caption{\sl ---Continued.}
  \begin{tabular}{lrrrcrrrcrrr} \\ \hline
   & \multicolumn{11}{c}{\sc Galaxy}\\              \cline{2-12} 
  &
 \multicolumn{3}{c}{J 16241012-00220258 }&&
 \multicolumn{3}{c}{UM 462e             }&&
 \multicolumn{3}{c}{J 01345200-00385440 } \\ 
  \cline{2-4} \cline{6-8} \cline{10-12}
  Ion
  &$F$($\lambda$)/$F$(H$\beta$)
  &$I$($\lambda$)/$I$(H$\beta$)
  &EW$^a$&
  &$F$($\lambda$)/$F$(H$\beta$)
  &$I$($\lambda$)/$I$(H$\beta$)
  &EW$^a$&
  &$F$($\lambda$)/$F$(H$\beta$)
  &$I$($\lambda$)/$I$(H$\beta$)
  &EW$^a$ \\ \hline
3727 [O {\sc ii}]                 & 135.28 $\pm$   1.97 & 149.63 $\pm$   2.31 & 101.7 & & 241.24 $\pm$   4.16 & 241.24 $\pm$   4.39 & 171.8 & & 105.63 $\pm$   4.39 & 115.80 $\pm$   5.08 & 115.2 \\
3750 H12                          &   2.02 $\pm$   0.16 &   3.79 $\pm$   0.39 &   1.6 & &   5.43 $\pm$   0.71 &   5.43 $\pm$   0.86 &   4.3 & &   ...~~~~~~ &   ...~~~~~~ &  ...~~~\\
3771 H11                          &   2.49 $\pm$   0.16 &   4.29 $\pm$   0.35 &   2.0 & &   6.14 $\pm$   0.76 &   6.14 $\pm$   0.92 &   4.6 & &   ...~~~~~~ &   ...~~~~~~ &  ...~~~\\
3798 H10                          &   3.53 $\pm$   0.18 &   5.42 $\pm$   0.33 &   2.8 & &   4.61 $\pm$   0.72 &   4.61 $\pm$   0.88 &   3.4 & &   ...~~~~~~ &   ...~~~~~~ &  ...~~~\\
\hline
  \end{tabular}
  \end{table}
\clearpage

\setcounter{table}{2}

  \begin{table}
  \centering
  \caption{\sl ---Continued.}
  \begin{tabular}{lrrrcrrrcrrr} \\ \hline
   & \multicolumn{11}{c}{\sc Galaxy}\\              \cline{2-12} 
  &
 \multicolumn{3}{c}{Tol 65b             }&&
 \multicolumn{3}{c}{UM 462b             }&&
 \multicolumn{3}{c}{CGCG 007-025d       } \\ 
  \cline{2-4} \cline{6-8} \cline{10-12}
  Ion
  &$F$($\lambda$)/$F$(H$\beta$)
  &$I$($\lambda$)/$I$(H$\beta$)
  &EW$^a$&
  &$F$($\lambda$)/$F$(H$\beta$)
  &$I$($\lambda$)/$I$(H$\beta$)
  &EW$^a$&
  &$F$($\lambda$)/$F$(H$\beta$)
  &$I$($\lambda$)/$I$(H$\beta$)
  &EW$^a$ \\ \hline
3727 [O {\sc ii}]                 &  86.92 $\pm$   1.50 &  95.83 $\pm$   1.74 &  45.3 & & 130.42 $\pm$   1.95 & 149.50 $\pm$   2.38 &  73.1 & & 109.73 $\pm$   1.94 & 107.04 $\pm$   2.03 &  43.2 \\
3771 H11                          &   ...~~~~~~ &   ...~~~~~~ &  ...~~~& &   2.18 $\pm$   0.27 &   4.58 $\pm$   0.68 &   1.2 & &   ...~~~~~~ &   ...~~~~~~ &  ...~~~\\
3798 H10                          &   4.61 $\pm$   0.51 &   5.97 $\pm$   0.87 &   2.7 & &   2.97 $\pm$   0.25 &   5.53 $\pm$   0.58 &   1.6 & &   ...~~~~~~ &   ...~~~~~~ &  ...~~~\\
3835 H9                           &   5.90 $\pm$   0.42 &   7.40 $\pm$   0.79 &   3.3 & &   4.62 $\pm$   0.26 &   7.40 $\pm$   0.52 &   2.5 & &   4.11 $\pm$   0.58 &   8.13 $\pm$   1.42 &   1.7 \\
\hline
  \end{tabular}
  \end{table}
\clearpage

\setcounter{table}{2}

  \begin{table}
  \centering
  \caption{\sl ---Continued.}
  \begin{tabular}{lrrrcrrrcrrr} \\ \hline
   & \multicolumn{11}{c}{\sc Galaxy}\\              \cline{2-12} 
  &
 \multicolumn{3}{c}{ESO 338-IG04        }&&
 \multicolumn{3}{c}{Mrk 1236b           }&&
 \multicolumn{3}{c}{IC 4662d            } \\ 
  \cline{2-4} \cline{6-8} \cline{10-12}
  Ion
  &$F$($\lambda$)/$F$(H$\beta$)
  &$I$($\lambda$)/$I$(H$\beta$)
  &EW$^a$&
  &$F$($\lambda$)/$F$(H$\beta$)
  &$I$($\lambda$)/$I$(H$\beta$)
  &EW$^a$&
  &$F$($\lambda$)/$F$(H$\beta$)
  &$I$($\lambda$)/$I$(H$\beta$)
  &EW$^a$ \\ \hline
3727 [O {\sc ii}]                 & 128.77 $\pm$   1.89 & 128.68 $\pm$   2.03 &  43.9 & & 156.56 $\pm$   2.38 & 159.86 $\pm$   2.64 &  55.6 & &  98.91 $\pm$   1.50 & 107.42 $\pm$   1.74 &  25.8 \\
3798 H10                          &   2.67 $\pm$   0.25 &   5.82 $\pm$   0.70 &   0.9 & &   2.14 $\pm$   0.47 &   6.88 $\pm$   1.79 &   0.8 & &   2.35 $\pm$   0.34 &   5.55 $\pm$   0.95 &   0.6 \\
3835 H9                           &   5.11 $\pm$   0.28 &   8.08 $\pm$   0.56 &   1.9 & &   3.16 $\pm$   0.34 &   7.88 $\pm$   1.06 &   1.1 & &   3.33 $\pm$   0.31 &   6.57 $\pm$   0.71 &   0.9 \\
3868 [Ne {\sc iii}]               &  47.48 $\pm$   0.73 &  47.33 $\pm$   0.76 &  16.6 & &  29.98 $\pm$   0.57 &  30.40 $\pm$   0.61 &  10.1 & &  39.40 $\pm$   0.64 &  42.20 $\pm$   0.72 &   9.9 \\
\hline
  \end{tabular}
  \end{table}
\clearpage

\setcounter{table}{2}

  \begin{table}
  \centering
  \caption{\sl ---Continued.}
  \begin{tabular}{lrrrcrrr} \\ \hline
   & \multicolumn{7}{c}{\sc Galaxy}\\              \cline{2-8} 
  &
 \multicolumn{3}{c}{J 01074656+01035206 }&&
 \multicolumn{3}{c}{IC 4662e            } \\
  \cline{2-4} \cline{6-8}
  Ion
  &$F$($\lambda$)/$F$(H$\beta$)
  &$I$($\lambda$)/$I$(H$\beta$)
  &EW$^a$&
  &$F$($\lambda$)/$F$(H$\beta$)
  &$I$($\lambda$)/$I$(H$\beta$)
  &EW$^a$ \\ \hline
3727 [O {\sc ii}]                 & 258.94 $\pm$   7.07 & 270.12 $\pm$   8.16 &  97.2 & & 338.12 $\pm$   9.73 & 324.66 $\pm$  10.85 &  28.5 \\
3868 [Ne {\sc iii}]               &  17.31 $\pm$   2.96 &  17.84 $\pm$   3.20 &   5.4 & &  74.31 $\pm$   4.21 &  70.97 $\pm$   4.42 &   4.9 \\
3968 [Ne {\sc iii}] + H7          &   ...~~~~~~ &   ...~~~~~~ &  ...~~~& &  12.31 $\pm$   3.11 &  20.60 $\pm$   6.21 &   1.0 \\
4101 H$\delta$                    &  23.99 $\pm$   2.12 &  28.93 $\pm$   3.24 &   9.1 & &  14.28 $\pm$   2.06 &  22.68 $\pm$   4.33 &   1.2 \\
4340 H$\gamma$                    &  42.75 $\pm$   2.01 &  47.23 $\pm$   2.96 &  15.6 & &  41.02 $\pm$   2.40 &  46.73 $\pm$   3.59 &   3.8 \\
4363 [O {\sc iii}]                &   4.20 $\pm$   1.26 &   4.17 $\pm$   1.31 &   1.4 & &   ...~~~~~~ &   ...~~~~~~ &  ...~~~\\
4861 H$\beta$                     & 100.00 $\pm$   2.86 & 100.00 $\pm$   3.39 &  40.0 & & 100.00 $\pm$   3.35 & 100.00 $\pm$   4.07 &   9.7 \\
4959 [O {\sc iii}]                &  69.59 $\pm$   2.29 &  66.26 $\pm$   2.28 &  28.6 & & 162.33 $\pm$   4.88 & 149.48 $\pm$   4.87 &  13.8 \\
5007 [O {\sc iii}]                & 206.84 $\pm$   5.10 & 196.34 $\pm$   5.06 &  84.6 & & 481.39 $\pm$  12.72 & 442.66 $\pm$  12.69 &  41.5 \\
 $C$(H$\beta$) & \multicolumn{3}{c}{ 0.120 }& & \multicolumn{3}{c}{ 0.085 } \\
 $F$(H$\beta$)$^b$ & \multicolumn{3}{c}{ 10.52 }& & \multicolumn{3}{c}{ 24.45 } \\
 EW(abs) \AA & \multicolumn{3}{c}{ 1.75 }& & \multicolumn{3}{c}{ 0.80 } \\ \hline

 $^a$ {in \AA} \\
 $^b$ {in units of 10$^{-16}$ ergs s$^{-1}$cm$^{-2}$}
  \end{tabular}
  \end{table}

\clearpage

\setcounter{table}{3}

  \begin{table}
\centering
  \caption{Ionic and Total Heavy Element Abundances\label{t3_1}} 
  \begin{tabular}{lrrrrrrrr} \\ \hline
  {}&\multicolumn{8}{c}{\sc Galaxy}\\ \cline{2-9}
  {Property}&
\multicolumn{1}{c}{Pox 186             } & \multicolumn{1}{c}{HS 2236+1344a       } &
\multicolumn{1}{c}{UM 461a             } & \multicolumn{1}{c}{Tol 1214-277        } &
\multicolumn{1}{c}{UM 570              } & \multicolumn{1}{c}{CGCG 032-017         } &
\multicolumn{1}{c}{Pox 108             } & \multicolumn{1}{c}{UM 559              } \\ \hline
  $T_{\rm e}$(O {\sc iii}) (K) &
$16736  \pm 169 $ & $20506  \pm 250 $ &
$16614  \pm 158 $ & $20421  \pm 262 $ &
$18308  \pm 216 $ & $13817  \pm 114 $ &
$13060  \pm 146 $ & $16052  \pm 177 $  \\ 

  $T_{\rm e}$(O {\sc ii}) (K) &
$14841  \pm 140 $ & $15634  \pm 175 $ &
$14792  \pm 131 $ & $15632  \pm 185 $ &
$15344  \pm 169 $ & $13255  \pm 102 $ &
$12705  \pm 134 $ & $14546  \pm 149 $  \\ 

  $T_{\rm e}$(S {\sc iii}) (K) &
$15684  \pm 140 $ & $18723  \pm 208 $ &
$15553  \pm 131 $ & $18690  \pm 217 $ &
$16859  \pm 179 $ & $12611  \pm  95 $ &
$12334  \pm 122 $ & $15268  \pm 147 $  \\ 

  \\ 
  O$^+$/H$^+$ ($\times$10$^4$) &
$ 0.033 \pm 0.001$ & $ 0.023 \pm 0.001$ &
$ 0.039 \pm 0.001$ & $ 0.021 \pm 0.001$ &
$ 0.018 \pm 0.001$ & $ 0.109 \pm 0.003$ &
$ 0.133 \pm 0.005$ & $ 0.086 \pm 0.003$ \\
\hline
\end{tabular}



  \begin{tabular}{lrrrrrrrr} \\ \hline
  {}&\multicolumn{8}{c}{\sc Galaxy}\\ \cline{2-9}
  {Property}&
\multicolumn{1}{c}{UM 442              } & \multicolumn{1}{c}{Mrk 1329            } &
\multicolumn{1}{c}{Mrk 1236a           } & \multicolumn{1}{c}{Tol 2146-391        } &
\multicolumn{1}{c}{CGCG 007-025a        } & \multicolumn{1}{c}{HS 2134+0400        } &
\multicolumn{1}{c}{J 0014-0043 } & \multicolumn{1}{c}{J 2324-0006 } \\ \hline
  $T_{\rm e}$(O {\sc iii}) (K) &
$12949  \pm 125 $ & $10467  \pm  70 $ &
$12141  \pm 108 $ & $16262  \pm 171 $ &
$15836  \pm 149 $ & $20226  \pm 392 $ &
$13911  \pm 381 $ & $14097  \pm 145 $  \\ 

  $T_{\rm e}$(O {\sc ii}) (K) &
$12619  \pm 114 $ & $10432  \pm  67 $ &
$11960  \pm 101 $ & $14641  \pm 143 $ &
$14443  \pm 127 $ & $15624  \pm 280 $ &
$13319  \pm 341 $ & $13443  \pm 129 $  \\ 

  $T_{\rm e}$(S {\sc iii}) (K) &
$12302  \pm 103 $ & $10448  \pm  58 $ &
$12184  \pm  90 $ & $15217  \pm 142 $ &
$14784  \pm 124 $ & $18695  \pm 326 $ &
$12940  \pm 316 $ & $13116  \pm 120 $  \\ 

  \\ 
  O$^+$/H$^+$ ($\times$10$^4$) &
$ 0.115 \pm 0.004$ & $ 0.329 \pm 0.009$ &
$ 0.256 \pm 0.008$ & $ 0.073 \pm 0.002$ &
$ 0.090 \pm 0.003$ & $ 0.037 \pm 0.002$ &
$ 0.138 \pm 0.011$ & $ 0.186 \pm 0.006$ \\
\hline
\end{tabular}
\end{table}

\clearpage

\setcounter{table}{3}

  \begin{table}
\centering
  \caption{\sl ---Continued.} 
  \begin{tabular}{lrrrrrrrr} \\ \hline
  {}&\multicolumn{8}{c}{\sc Galaxy}\\ \cline{2-9}
  {Property}&
\multicolumn{1}{c}{J 1253-0312 } & \multicolumn{1}{c}{Tol 65a             } &
\multicolumn{1}{c}{UM 311               } & \multicolumn{1}{c}{J 1304-0333 } &
\multicolumn{1}{c}{J 0248-0817a} & \multicolumn{1}{c}{Pox 4a              } &
\multicolumn{1}{c}{Pox 120a            } & \multicolumn{1}{c}{HS 2236+1344b       } \\ \hline
  $T_{\rm e}$(O {\sc iii}) (K) &
$13821  \pm 103 $ & $17649  \pm 209 $ &
$ 9910  \pm 113 $ & $10754  \pm 222 $ &
$13158  \pm 103 $ & $13939  \pm 110 $ &
$15259  \pm 136 $ & $18648  \pm 237 $  \\ 

  $T_{\rm e}$(O {\sc ii}) (K) &
$13257  \pm  92 $ & $15163  \pm 167 $ &
$ 9895  \pm 109 $ & $10777  \pm 212 $ &
$12779  \pm  94 $ & $13338  \pm  99 $ &
$14145  \pm 117 $ & $15420  \pm 182 $  \\ 

  $T_{\rm e}$(S {\sc iii}) (K) &
$12803  \pm  85 $ & $16935  \pm 174 $ &
$ 9886  \pm  94 $ & $10777  \pm 184 $ &
$12414  \pm  86 $ & $12867  \pm  92 $ &
$14088  \pm 113 $ & $17592  \pm 197 $  \\ 

  \\ 
  O$^+$/H$^+$ ($\times$10$^4$) &
$ 0.121 \pm 0.003$ & $ 0.060 \pm 0.002$ &
$ 0.733 \pm 0.034$ & $ 0.550 \pm 0.040$ &
$ 0.133 \pm 0.004$ & $ 0.100 \pm 0.003$ &
$ 0.072 \pm 0.002$ & $ 0.039 \pm 0.001$ \\
\hline
\end{tabular}



  \begin{tabular}{lrrrrrrrr} \\ \hline
  {}&\multicolumn{8}{c}{\sc Galaxy}\\ \cline{2-9}
  {Property}&
\multicolumn{1}{c}{CGCG 007-025c        } & \multicolumn{1}{c}{UM 462a             } &
\multicolumn{1}{c}{Tol 2138-405        } & \multicolumn{1}{c}{Pox 4b              } &
\multicolumn{1}{c}{IC 4662a            } & \multicolumn{1}{c}{J 1430+0027 } &
\multicolumn{1}{c}{IC 4662c            } & \multicolumn{1}{c}{Pox 120b            } \\ \hline
  $T_{\rm e}$(O {\sc iii}) (K) &
$15575  \pm 248 $ & $13932  \pm 136 $ &
$13899  \pm 165 $ & $14336  \pm  75 $ &
$12731  \pm 103 $ & $13027  \pm 138 $ &
$12459  \pm 112 $ & $13695  \pm 147 $  \\ 

  $T_{\rm e}$(O {\sc ii}) (K) &
$14312  \pm 212 $ & $13333  \pm 122 $ &
$13311  \pm 148 $ & $13599  \pm  87 $ &
$12448  \pm  94 $ & $12679  \pm 126 $ &
$12227  \pm 104 $ & $13170  \pm 132 $  \\ 

  $T_{\rm e}$(S {\sc iii}) (K) &
$14704  \pm 206 $ & $12888  \pm 113 $ &
$12887  \pm 137 $ & $13451  \pm  63 $ &
$12216  \pm  85 $ & $12338  \pm 115 $ &
$12180  \pm  93 $ & $12780  \pm 122 $  \\ 

  \\ 
  O$^+$/H$^+$ ($\times$10$^4$) &
$ 0.120 \pm 0.005$ & $ 0.169 \pm 0.005$ &
$ 0.192 \pm 0.007$ & $ 0.240 \pm 0.007$ &
$ 0.247 \pm 0.007$ & $ 0.222 \pm 0.008$ &
$ 0.296 \pm 0.009$ & $ 0.172 \pm 0.006$ \\
\hline
\end{tabular}
\end{table}

\clearpage

\setcounter{table}{3}

  \begin{table}
\centering
  \caption{\sl ---Continued.} 
  \begin{tabular}{lrrrrrrrr} \\ \hline
  {}&\multicolumn{8}{c}{\sc Galaxy}\\ \cline{2-9}
  {Property}&
\multicolumn{1}{c}{g252292-171050     } & \multicolumn{1}{c}{UM 461b             } &
\multicolumn{1}{c}{Mrk 1236c           } & \multicolumn{1}{c}{J 1448-0110 } &
\multicolumn{1}{c}{IC 4662b            } & \multicolumn{1}{c}{Pox 105             } &
\multicolumn{1}{c}{UM 462d             } & \multicolumn{1}{c}{UM 462c             } \\ \hline
  $T_{\rm e}$(O {\sc iii}) (K) &
$15870  \pm 246 $ & $15786  \pm 226 $ &
$12249  \pm 349 $ & $13877  \pm 134 $ &
$12732  \pm 103 $ & $15270  \pm 205 $ &
$14529  \pm 278 $ & $14768  \pm 175 $  \\ 

  $T_{\rm e}$(O {\sc ii}) (K) &
$14459  \pm 209 $ & $14418  \pm 193 $ &
$12052  \pm 324 $ & $13296  \pm 120 $ &
$12449  \pm  95 $ & $14151  \pm 177 $ &
$13720  \pm 245 $ & $13865  \pm 153 $  \\ 

  $T_{\rm e}$(S {\sc iii}) (K) &
$14795  \pm 205 $ & $14892  \pm 188 $ &
$11951  \pm 290 $ & $12870  \pm 111 $ &
$12216  \pm  86 $ & $14214  \pm 170 $ &
$13637  \pm 231 $ & $13764  \pm 145 $  \\ 

  \\ 
  O$^+$/H$^+$ ($\times$10$^4$) &
$ 0.071 \pm 0.003$ & $ 0.073 \pm 0.003$ &
$ 0.459 \pm 0.040$ & $ 0.130 \pm 0.004$ &
$ 0.130 \pm 0.004$ & $ 0.120 \pm 0.005$ &
$ 0.265 \pm 0.014$ & $ 0.223 \pm 0.008$ \\
\hline
\end{tabular}



  \begin{tabular}{lrrrrrrrr} \\ \hline
  {}&\multicolumn{8}{c}{\sc Galaxy}\\ \cline{2-9}
  {Property}&
\multicolumn{1}{c}{J 0248-0817b} & \multicolumn{1}{c}{CGCG 007-025b        } &
\multicolumn{1}{c}{J 1624-0022 } & \multicolumn{1}{c}{UM 462e             } &
\multicolumn{1}{c}{J 0134-0038 } & \multicolumn{1}{c}{Tol 65b             } &
\multicolumn{1}{c}{UM 462b             } & \multicolumn{1}{c}{CGCG 007-025d        } \\ \hline
  $T_{\rm e}$(O {\sc iii}) (K) &
$13535  \pm 376 $ & $15850  \pm 200 $ &
$12021  \pm 120 $ & $15873  \pm 478 $ &
$14894  \pm1469 $ & $16910  \pm 361 $ &
$14152  \pm 182 $ & $16762  \pm 388 $  \\ 

  $T_{\rm e}$(O {\sc ii}) (K) &
$13057  \pm 339 $ & $14449  \pm 170 $ &
$11857  \pm 112 $ & $14461  \pm 406 $ &
$13939  \pm1282 $ & $14909  \pm 296 $ &
$13480  \pm 162 $ & $14852  \pm 320 $  \\ 

  $T_{\rm e}$(S {\sc iii}) (K) &
$12676  \pm 312 $ & $15013  \pm 166 $ &
$12142  \pm 100 $ & $14966  \pm 397 $ &
$13943  \pm1219 $ & $16355  \pm 299 $ &
$13218  \pm 151 $ & $16102  \pm 322 $  \\ 

  \\ 
  O$^+$/H$^+$ ($\times$10$^4$) &
$ 0.172 \pm 0.014$ & $ 0.125 \pm 0.004$ &
$ 0.302 \pm 0.010$ & $ 0.245 \pm 0.019$ &
$ 0.133 \pm 0.035$ & $ 0.089 \pm 0.005$ &
$ 0.191 \pm 0.007$ & $ 0.100 \pm 0.006$ \\
\hline
\end{tabular}
\end{table}

\clearpage

\setcounter{table}{3}

  \begin{table}
\centering
  \caption{\sl ---Continued.} 
  \begin{tabular}{lrrrrr} \\ \hline
  {}&\multicolumn{5}{c}{\sc Galaxy}\\ \cline{2-6}
  {Property}&
\multicolumn{1}{c}{ESO 338-IG04        } & \multicolumn{1}{c}{Mrk 1236b           } &
\multicolumn{1}{c}{IC 4662d            } & \multicolumn{1}{c}{J 0107+0103 } &
\multicolumn{1}{c}{IC 4662e            } \\ \hline
  $T_{\rm e}$(O {\sc iii}) (K) &
$14634  \pm 164 $ & $11694  \pm 232 $ &
$12522  \pm 170 $ & $15639  \pm2330 $ &
$13089  \pm 113 $  \\ 

  $T_{\rm e}$(O {\sc ii}) (K) &
$13784  \pm 144 $ & $11568  \pm 217 $ &
$12279  \pm 157 $ & $14344  \pm1993 $ &
$12727  \pm 128 $  \\ 

  $T_{\rm e}$(S {\sc iii}) (K) &
$13601  \pm 136 $ & $11724  \pm 193 $ &
$12123  \pm 142 $ & $15055  \pm1934 $ &
$12359  \pm  94 $  \\ 

  \\ 
  O$^+$/H$^+$ ($\times$10$^4$) &
$ 0.153 \pm 0.005$ & $ 0.354 \pm 0.023$ &
$ 0.191 \pm 0.008$ & $ 0.282 \pm 0.107$ &
$ 0.507 \pm 0.023$  \\
\hline
\end{tabular}
\end{table}
  \end{landscape}

\end{document}